\documentclass[a4paper,11pt]{article}
\pdfoutput=1
\usepackage{jcappub}
\usepackage{amsmath, amssymb, graphics, epsfig}
\usepackage{graphicx}
\usepackage{enumerate}
\usepackage{color}
\usepackage{soul}
\usepackage{multirow}
\usepackage{booktabs}

\usepackage[normalem]{ulem}

\usepackage{xspace}
\newcommand{\AP}{Alcock-Paczy\'{n}ski\xspace}

\newcommand{\hmpc}{$h^{-1}\mathrm{Mpc}$\xspace}
\newcommand{\hmsun}{$h^{-1}M_\odot$\xspace}
\newcommand{\lcdm}{$\Lambda$CDM\xspace}
\newcommand{\vide}{\tt VIDE\normalfont\xspace}
\newcommand{\zobov}{\tt ZOBOV\normalfont\xspace}

\title{\boldmath{\fontsize{20}{11}\selectfont {\centering DEMNUni: disentangling dark energy from massive neutrinos with the void size function}}}

\author[a,b]{Giovanni Verza,}
\author[c]{Carmelita Carbone,}
\author[d,e,f]{Alice Pisani,}
\author[b,a]{and Alessandro Renzi}

\affiliation[a]{INFN, Sezione di Padova,
via Marzolo 8, I-35131 Padova, Italy}
\affiliation[b]{Dipartimento di Fisica e Astronomia ``G. Galilei",
Universit\`a degli Studi di Padova, via Marzolo 8, I-35131 Padova, Italy}
\affiliation[c]{INAF – Istituto di Astrofisica Spaziale e Fisica cosmica di Milano (IASF-MI), Via Alfonso Corti 12, I-20133 Milano, Italy}
\affiliation[d]{The Cooper Union for the Advancement of Science and Art, 41 Cooper Square, New York, NY 10003, USA}
\affiliation[e]{Center for Computational Astrophysics, Flatiron Institute, 162 5$^{\rm th}$ Avenue, 10010, New York, NY, USA}
\affiliation[f]{Department of Astrophysical Sciences, Peyton Hall, Princeton University, Princeton, NJ 08544, USA}

\emailAdd{giovanni.verza@nyu.edu}
\emailAdd{carmelita.carbone@inaf.it}
\emailAdd{apisani@astro.princeton.edu}
\emailAdd{alessandro.renzi@pd.infn.it}

\abstract{
Cosmic voids, the underdense regions in the Universe, are impacted by dark energy and massive neutrinos. In this work, relying on the DEMNUni suite of cosmological simulations, we explore the void size function in cosmologies with both dynamical dark energy and massive neutrinos. We investigate the impact of different choices of dark matter tracers on the void size function and study its sensitivity to the joint effect of several dark energy equations of state and total neutrino masses. 
In particular, we find that for all the combinations of neutrino mass and dark energy equation of state considered, the differences between the corresponding void size functions are larger than the associated Poisson noise, and therefore can be all distinguished. This looks very promising considering that forthcoming surveys, as the Euclid satellite, will be characterised by a void statistics with similar number densities and volumes as for the DEMNUni suite. These findings show that the use of the void size function in forthcoming large galaxy surveys will be extremely useful in breaking degeneracies among these cosmological parameters.
}

\begin{document}

\maketitle

\section{Introduction}

Cosmic voids are extended underdense regions spanning a wide range of scales, and represent the largest observable objects in the Universe. Their size and underdense nature make them particularly suited to probe dark energy (DE), modified gravity~\cite{lee_2009,lavaux_wandelt_2009,biswas_2010,li_2010_MG,clampitt_2013_MG,spolyar_2013,cai_2015,pisani_2015_abundance,pollina_2015,zivick_2015,achitouv_2016,sahlen_2016,falck_2018,sahlen_2018,paillas_2019,perico_2019,verza_2019,contarini_2021,euclid_vsf,wilson_2022_voids_MG}, and massive neutrinos~\cite{massara_2015,banerjee_2016,kreisch_2019,sahlen_2019,schuster_2019,zhang_2020, kreisch_2021,euclid_vsf}. Many cosmological statistics of cosmic voids have been recently explored, such as the void size function (VSF) and the void-galaxy cross-correlation, which can probe the underlying cosmological model of the Universe~\citep{pisani_2015_abundance,hamaus_2015,hamaus_2016,sahlen_2016,cai_2016,chuang_2017,achitouv_2017,hawken_2017,hamaus_2017,sahlen_2018,achitouv_2019,nadathur_2019_BOSS,kreisch_2019,contarini_2019,verza_2019,pisani_2019,hamaus_2020,hawken_2020,nadathur_2020,aubert_2020,bayer_2021,kreisch_2021,moresco_2022_cosmological_probes,euclid_vsf,schuster_2022,contarini_2022_sdss,contarini_2022_tensions,radinovic_2023}. 
Cosmic voids are large extended objects, therefore their usage for cosmological analyses requires galaxy surveys of large volume, but also deep enough to map in detail contiguous regions of the observable Universe~\citep{hamaus_2022_euclid,euclid_vsf}.
Ongoing and upcoming spectroscopic and photometric galaxy surveys, such as BOSS~\cite{dawson_2013_BOSS}, DES~\cite{DES_2016}, DESI~\cite{DESI_2016}, PFS~\cite{PFS_2016}, Euclid~\cite{laureijs_20211_euclid_report}, the Roman Space Telescope~\cite{spergel_2015_WFIRST}, SPHEREx~\cite{dore_2018_SPHEREx}, and the Vera Rubin Observatory~\cite{ivezic_2019_LSST}, fulfil these requirements, bringing cosmic void statistics among the new competitive cosmological probes~\citep{pisani_2019,moresco_2022_cosmological_probes}. 
In this work we investigate the properties of the VSF, i.e. the number density of voids as a function of their size, and its sensitivity to the DE equation of state (EoS) and neutrino mass, using a suite of large cosmological simulations which share the same amplitude of primordial scalar perturbations, $A_s$, as measured by Planck\footnote{The choice of a fixed $A_s$ implies different values of $\sigma_8$ at $z=0$, and may correlate void statistics with CMB anisotropies. However, the choice of a fixed $\sigma_8$, obtained as a derived parameter from CMB experiments, may implicitly impose a further prior on the cosmological model with which the $\sigma_8$ value is derived, e.g. the \lcdm model. We discuss about these different choices in the Appendix, showing that they do not impact the results presented in this work.}.
DE and massive neutrinos can produce degenerate effects on cosmological observables especially if not combined with CMB priors~\citep{hamann_2012,wang_2012,santos_2013,zhang_2014,zablocki_2016,wang_2016,lorenz_2017,zhao_2017,li_2018,mishra-sharma_2018,upadhye_2019,diaz_rivero_2019,zhao_2020,zhang_Mnu_2020}. In this work, we found that the VSF is particularly able to mitigate the existing degeneracy between DE and the total neutrino mass. This finding shows the importance of cosmic voids in cosmological analyses, possibly completing the cosmological information carried by other cosmological probes, such as the statistics of the observed CMB anisotropies, galaxy clusters, galaxy clustering, and weak lensing.

Voids are detectable as underdensities in the distribution of tracers of the underlying dark matter density field, therefore in the first part of this work we explore the impact on the VSF of various tracers, i.e. haloes with different masses and cold dark matter (CDM) particles. We then thoroughly investigate the sensitivity of the VSF to DE-EoS, total neutrino mass and different combination of both the DE-EoS and the sum of neutrino masses, finding that the VSF can distinguish among all the combinations explored in this work.

\section{Simulations and void finder}

For this study, we use the “Dark Energy and Massive Neutrino Universe” (DEMNUni) suite of large N-body simulations~\cite{carbone_2016_demnuni}. The DEMNUni simulations have been produced with the aim of investigating the large-scale structure of the Universe in the presence of massive neutrinos and dynamical DE, and they were conceived for the nonlinear analysis and modelling of different probes, including dark matter, halo, and galaxy clustering~\cite{castorina_2015,moresco_2016,zennaro_2018,ruggeri_2018,bel_2019,parimbelli_2021,parimbelli_2022,Guidi_2022,Baratta_2022,SHAM-Carella_in_prep}, weak lensing, CMB lensing, Sunyaev-Zel'dovich and integrated Sachs-Wolfe effects~\cite{carbone_2016_demnuni,roncarelli_2015,fabbian_2018,Beatriz_2023}, cosmic void statistics~\cite{kreisch_2019,schuster_2019,verza_2019,verza_2022}, as well as cross-correlations among these probes~\cite{Vielzeuf2023,Cuozzo2022}.
The DEMNUni simulations combine a good mass resolution with a large volume to include perturbations both at large and small scales. They are characterised by a comoving volume of $(2 \: h^{-1}\mathrm{Gpc})^3$ filled with $2048^3$ dark matter particles and, when present, $2048^3$ neutrino particles. The simulations are initialised at $z_{\rm in}=99$ with Zel'dovich initial conditions. The initial power spectrum is rescaled to the initial redshift via the rescaling method developed in~\cite{zennaro_2017}. Initial conditions are then generated with a modified version of the \texttt{N-GenIC} software, assuming Rayleigh random amplitudes and uniform random phases.
The DEMNUni set is composed by 15 simulations, implementing the cosmological constant and 4 dynamical DE-EoS for each of the total neutrino masses considered in the degenerate mass scenario with three active neutrinos, i.e. $\sum m_\nu = 0,\, 0.16,\, 0.32\, {\rm eV}$.
The four DE-EoS variants are parameterised via the Chevallier–Polarski–Linder (CPL) parameterisation~\cite{chevallier_2001,linder_2003}
\begin{figure}[t!]
\centering
\includegraphics[width=0.95\textwidth]{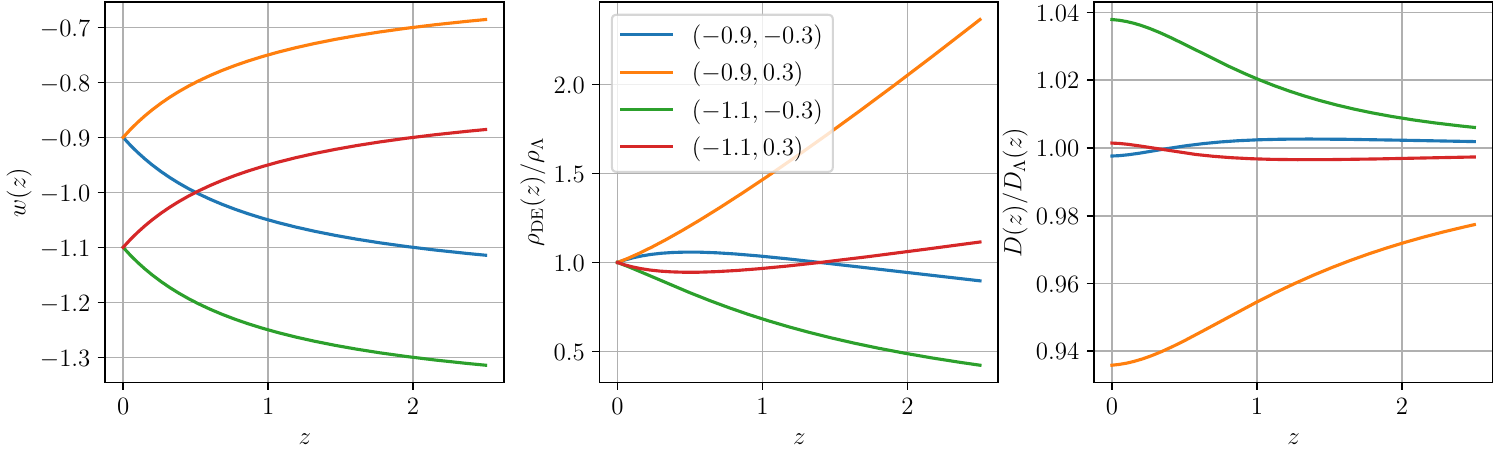}
\caption{DE-EoS (left panel), DE density with respect to the cosmological constant density (central panel), linear growth factor in the massless neutrino case for the considered DE-EoS with respect to \lcdm (right panel). The colour labels the DE-EoS as: 
$(w_0=-0.9,w_{\rm a}=-0.3)$ (blue), $(w_0=-0.9,w_{\rm a}=0.3)$ (orange), $(w_0=-1.1,w_{\rm a}=-0.3)$ (green), and $(w_0=-1.1,w_{\rm a}=0.3)$ (red).}
\label{fig:DE_EoS}
\end{figure}
\begin{equation}
w(a) = w_0 + (1-a) \, w_{\rm a} \quad \Rightarrow \quad w(z) = w_0 + \frac{z}{1+z} w_{\rm a}\, ,
\end{equation}
where the parameters $(w_0,w_{\rm a})$ are $(-1,0)$ for the cosmological constant case and the 4 combinations between $w_0=[-0.9, \: -1.1]$ and $w_{\rm a} = [-0.3, \: 0.3]$ for dynamical DE. Fig.~\ref{fig:DE_EoS} shows some of the main cosmological quantities related to the DE-EoS implemented in the DEMNUni simulations, which are: the EoS (left panel); the ratio of the DE density $\rho_{\rm DE}$ to the cosmological constant energy density, i.e. $\rho_\Lambda = - \Lambda / (8 \pi G)$ (middle panel); the linear growth factor of perturbations in the presence of dynamical DE with respect to the \lcdm one for the massless neutrino case (right panel).
The other cosmological parameters of the simulations are based on a Planck 2013~\cite{planck2013} \lcdm reference cosmology (with massless neutrinos), in particular: $h=0.67$, $n_{\rm s}=0.96$, $A_{\rm s}=2.1265 \times 10^{-9}$, $\Omega_{\rm b}=0.05$, and $\Omega_{\rm m}=\Omega_{\rm CDM} + \Omega_{\rm b} + \Omega_\nu =0.32$; where $h=H_0/[100\, 
{\rm km} \, s^{-1}{\rm Mpc}^{-1}]$ and $H_0$ is the Hubble constant at the present time, $n_{\rm s}$ is the spectral index of the initial scalar perturbations, $A_{\rm s}$ is the scalar amplitude, $\Omega_{\rm b}$ the baryon density parameter, $\Omega_{\rm m}$ is the total matter density parameter, $\Omega_{\rm CDM}$ the cold dark matter density parameter, and $\Omega_\nu$ the neutrino density parameter. In the presence of massive neutrinos, $\Omega_{\rm b}$ and $\Omega_{\rm m}$ are kept fixed to the above values, while $\Omega_{\rm CDM}$ is changed accordingly. Tab.~\ref{tab:neutrino_params} summarises the masses of the CDM and neutrino particles together with the neutrino fraction $f_\nu \equiv \Omega_\nu / \Omega_{\rm m}$. 
\begin{table}[t]
\centering
\vspace{2ex}
\setlength{\tabcolsep}{0.7em}
\begin{tabular}{cccc}
\toprule
$\sum m_\nu$  [eV] & $f_\nu$ & $m_{\rm p}^{\rm CDM}$  [\hmsun] & $m_{\rm p}^\nu$  [\hmsun] \\
\midrule
\hline 
0     & 0 & $8.27\times 10^{10}$ & 0 \\
0.16  & 0.012 & $8.17\times 10^{10}$ & $0.99\times 10^9$ \\
0.32 & 0.024 & $8.07\times 10^{10}$ & $1.98\times 10^9$ \\
\noalign{\vspace{1ex}}
\hline 
\bottomrule
\end{tabular}
\caption{Summary of particle masses and neutrino fractions implemented in the DEMNUni simulations. The first column shows the total neutrino mass, the second the fraction of neutrinos and matter density parameters, and the last two columns show the corresponding mass of CDM and neutrino particles implemented in the simulations. 
}
\label{tab:neutrino_params}
\end{table}
Dark matter haloes are identified using a friends-of-friends (FoF) algorithm~\cite{davis_1985_fof} applied to dark matter particles, with a minimum number of particles fixed to 32, corresponding to a mass of $\sim 2.5 \times 10^{12} h^{-1}M_{\odot}$, and a linking length of 0.2 times the mean particle separation. FoF haloes are further processed with the {\sc subfind} algorithm~\cite{springel_2001_gadeget,dolang_2009_gadget} to produce subhalo catalogues. With this procedure, some of the initial FoF parent haloes are split into multiple substructures. In the following with the term ``halo'' we will refer to the objects identified by the {\sc subfind} algorithm.

To identify voids and build void catalogues, we use the second version of the ``Void IDentification and Examination'' (\href{http:www.cosmicvoids.net}{\vide}) public toolkit~\cite{sutter_2015_vide}.
\vide is based on the Voronoi tessellation~\cite{schaap_2000_dtfe} plus watershed void finding technique~\cite{platen_2007} implemented in \zobov~\cite{neyrinck_2008}.
The algorithm detects the minima of the density field and groups nearby Voronoi cells into zones, corresponding to local catchment ``basins", that are identified as voids.

For the present analysis, we consider the following redshift of comoving snapshots for each cosmology: $z=0$, 0.49, 1.05, 1.46, 2.05.
For each of them, we build catalogues of CDM-traced voids, corresponding to different dilutions of the CDM particle distribution, and four different catalogues of halo-traced voids corresponding to four minimum halo masses: $2.5 \times 10^{12}$\hmsun, $10^{13}$\hmsun, $2.5 \times 10^{13}$\hmsun, and $10^{14}$\hmsun.
Concerning CDM-traced voids, we run the void finder on various subsamples of the CDM particle catalogues: 
i)~we randomly diluted at 1.5\% the original dark matter particle distribution, ending up with $\sim 1.29 \times 10^{8}$ particles for each comoving snapshot; ii)~we randomly sampled the dark matter particle distribution to match the number density of haloes detected in the corresponding snapshot. We repeated this procedure for each redshift and halo mass-cut considered. In the following, with CDM-traced voids we indicate voids detected\footnote{We specify here that, in the case of massive neutrino particles, \vide detects voids using as tracers the CDM particles alone.} in the CDM distribution subsampled at 1.5\%; otherwise, we explicitly specify the kind of subsample considered.

We characterise the detected voids according to the void size, measured by \vide via the void effective radius, $R_{\rm eff}$, i.e. the radius of a sphere with the same volume of the void, $R_{\rm eff} = \left[ (3/4 \pi)  \sum_i V_i \right]^{1/3}$, with $V_i$ the volume of the $i^{\rm th}$ Voronoi cell building up the void.
\vide is a parameter-free algorithm and detects all the relative minima in the tracer distribution. Therefore, we prune the original void catalogue to consider only real underdensities, i.e. voids for which the mean tracer density within a sphere of radius $R_\mathrm{eff}/4$ is less than the mean tracer density of the comoving snapshot, for each of the tracers considered. 
We wish voids to be underdense with respect to the mean density, so we prune the original VIDE void catalogue considering voids for which the mean tracer density within a sphere of radius $R_\mathrm{eff}/4$ is less than the mean tracer density of the comoving snapshot, for each of the tracers considered.
\begin{table}[t!]
\centering
\begin{tabular}{ c | c c c c c }
\toprule
\hline 
\multirow{2}{*}{$z$} & \multirow{2}{*}{CDM} &\multicolumn{4}{c}{$M_{\rm min} \quad [h^{-1}M_\odot]$} \\
\cline{3-6}
& & $\ 2.5 \times 10^{12}\ $ & $\ 10^{13} \ $ & $\ 2.5 \times 10^{13} \ $ & $ \ 10^{14} \ $ \\
\hline 
0  & 4.0 & 7.9 & 12.3 & 17.3 & 31.5  \\
0.49  & 4.0 & 8.2 & 13.4 & 19.8 & 41.6  \\
1.05  & 4.0 & 8.9 & 15.7 & 25.5 & 67.5  \\
1.46  & 4.0 & 9.8 & 18.5 & 32.5 & 107.4 \\
2.05  & 4.0 & 11.8 & 25.2 & 51.3 & 260.6  \\
\hline 
\bottomrule
\end{tabular}
\caption{Mean tracer separation, in \hmpc units, in the \lcdm cosmology for the CDM and halo distributions associated to each of the minimum halo masses ($M_{\rm min}$) and redshifts considered.}
\label{tab:mts}
\end{table}

\section{Impact of tracers on the void size function}

\begin{figure}[t!]
\centering
\includegraphics[width=0.95\textwidth]{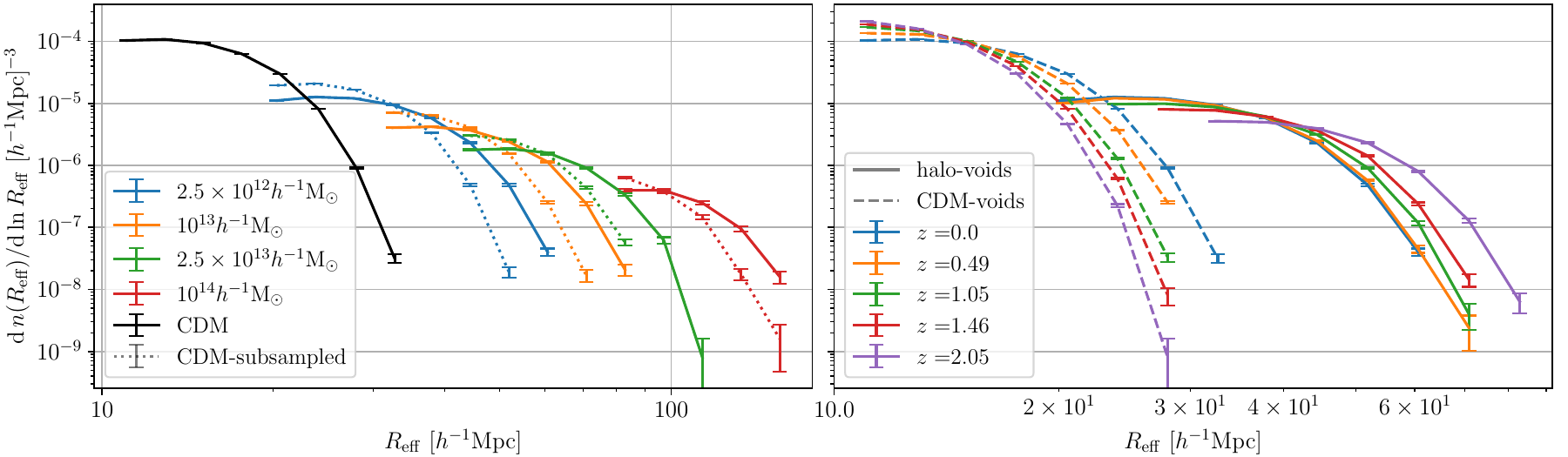}
\caption{Left: Solid lines show the VSF measured in the \lcdm simulation at $z=0$ for CDM-traced voids (black) and voids traced by haloes with $M>2.5 \times 10^{12}$\hmsun (blue), $10^{13}$\hmsun (orange), $2.5 \times 10^{13}$\hmsun (green), $10^{14}$\hmsun (red), respectively. With the same colours, dotted lines show the VSF for voids traced in the CDM distributions randomly subsampled to match the halo density for the corresponding halo mass-cuts. Right: redshift evolution of the VSF of CDM-traced voids (dashed lines) and halo-traced voids (solid lines) for $M \geq 2.5 \times 10^{12} h^{-1} M_\odot$ in the \lcdm simulation. The errorbars represent Poissonian errors.}
\label{fig:vide_voids_tracers_redshift_dep}
\end{figure}

The VSF is the number density of voids as a function of their size. According to the widely used notation, we express it as ${\rm d} n(R_{\rm eff}) / {\rm d}\ln R_{\rm eff}$, where $n(R_{\rm eff})$ is the number density of voids with radius $R_{\rm eff}$. Voids can be detected in any population of tracers of the underlying dark matter field. Voids detected in different tracer populations show different features, reflected in the corresponding VSF.

The left panel of Fig.~\ref{fig:vide_voids_tracers_redshift_dep} shows the VSF measured for the \lcdm cosmology in the CDM- and halo-traced void catalogues at $z=0$. In particular, solid lines with different colours represent CDM tracers subsampled at 1.5\% (black) and haloes tracers with different minimum masses: $2.5 \times 10^{12}$\hmsun (blue), $10^{13}$\hmsun (orange), $2.5 \times 10^{13}$\hmsun (green), and $10^{14}$\hmsun (red), respectively. Dotted lines show the VSF in the CDM particles distribution randomly subsampled to match the number density of haloes, the colours show the corresponding halo mass-cut. {The errorbars show the Poissonian error.
In estimating the density field via Voronoi tessellation and the derived watershed regions~\citep{neyrinck_2008}, each tracer population traces density fluctuations on different scales, therefore we expect different VSF curves for each different population. 
More massive haloes trace the underlying density field on larger scales with respect to less massive haloes, and therefore they probe underdensities (voids) on larger scales than less massive haloes. 
This is due to the bias and number density of tracers, which introduce a natural spatial smoothing scale via the mean tracer separation. The mean tracer separation is the typical distance between two objects of the same population (CDM or haloes), calculated as $(N_{\rm tr}/V)^{-1/3}$, where $N_{\rm tr}$ is the number of tracers in the comoving volume $V$. Tab.~\ref{tab:mts} lists the mean separation of the tracers in the \lcdm cosmology at the redshifts considered in this work. Even if in overdense regions it is possible to have a spatial resolution below the mean separation, due to the highly clustered objects, this is not true for underdensities, where the number density of objects is lower than the corresponding mean number density of the Universe and the clustering is suppressed. It follows that in underdense regions, the underlying matter fluctuations on a scale around or smaller than the mean tracer separation are beyond the spatial resolution and cannot be detected in the tracer distribution. Therefore, the detection of voids smaller than the mean tracer separation is related to numerical and/or Poissonian noise (due to tracer discreteness and sparsity) that on such small scales dominates over the physical signal~\citep{cousinou_2019}. The effect of tracer discreteness is reflected by the fact that both the halo-traced VSF and VSF in the corresponding random-sampled CDM distribution span approximately the same range of scales. On the other hand, tracer bias affects the amplitude of the VSF as we observe in the left panel of Fig.~\ref{fig:vide_voids_tracers_redshift_dep} that halo-traced large voids are more abundant than voids with comparable size in the corresponding subsampled CDM-void catalogue. This is due to the fact that the power spectrum of the halo field is biased, larger than the corresponding CDM one, entailing more abundant voids. The suppression of small halo-traced voids with respect to the subsampled CDM-traced ones is due to volume conservation and void merging: the higher number of large voids is due to the merging of smaller underdensities that in the biased tracer case are detected as one. Moreover, the sum of all void volumes cannot exceed the snapshot volume; therefore, when the number of large voids increases, the number of small voids decreases as a consequence.

The right panel of Fig.~\ref{fig:vide_voids_tracers_redshift_dep} shows the redshift evolution of the VSF of CDM-traced voids (dashed lines) and of halo-traced voids (solid lines) for $M \geq 2.5 \times 10^{12}$\hmsun in \lcdm cosmology. 
The redshift evolution of the void size function is impacted by two concurring effects: the evolution of matter density perturbations and the evolution of the population used as dark matter tracer.
On the one hand, CDM-traced voids are sensitive to the former effect only, since the total number of CDM particles is fixed and their bias is identically equal to 1. As expected, CDM-traced voids, i.e. underdense minima in the matter field, expand as the redshift decreases: for large voids, a fixed value of the VSF corresponds to larger effective void radii as time passes. 
Consequently, given the increase of volume for large voids, for a fixed mass resolution of the simulations the value of the CDM void size function decreases for small voids when redshift decreases. 
On the other hand, halo-traced voids are also impacted by the redshift evolution of the halo distribution. Haloes with a given minimum mass become more rare as the redshift increases, while the corresponding mean tracer separation and bias increase too. 
This means that for a fixed minimum halo mass, high-redshift haloes trace the underlying matter fluctuations on larger scales. As the redshift decreases, the number density of haloes increases; consequently, they trace the underlying matter distribution at smaller scales, and smaller voids can be detected. 
Effectively, this corresponds to smoothing the matter perturbation on a decreasing scale as time passes: at high redshifts density minima close to each other cannot be resolved as the smoothing scale is too large, and therefore they are detected as a single larger underdensity. Summarising, the right panel of Fig.~\ref{fig:vide_voids_tracers_redshift_dep} shows that the evolution of the halo population with fixed minimum mass drives the VSF of halo-traced voids toward larger radii at high redshifts and to smaller voids at low redshifts.

\section{Massive neutrinos and dark energy effects on the VSF}\label{sec:DE_MNU_VSF}
\begin{figure}[t!]
\centering
\includegraphics[width=0.95\textwidth]{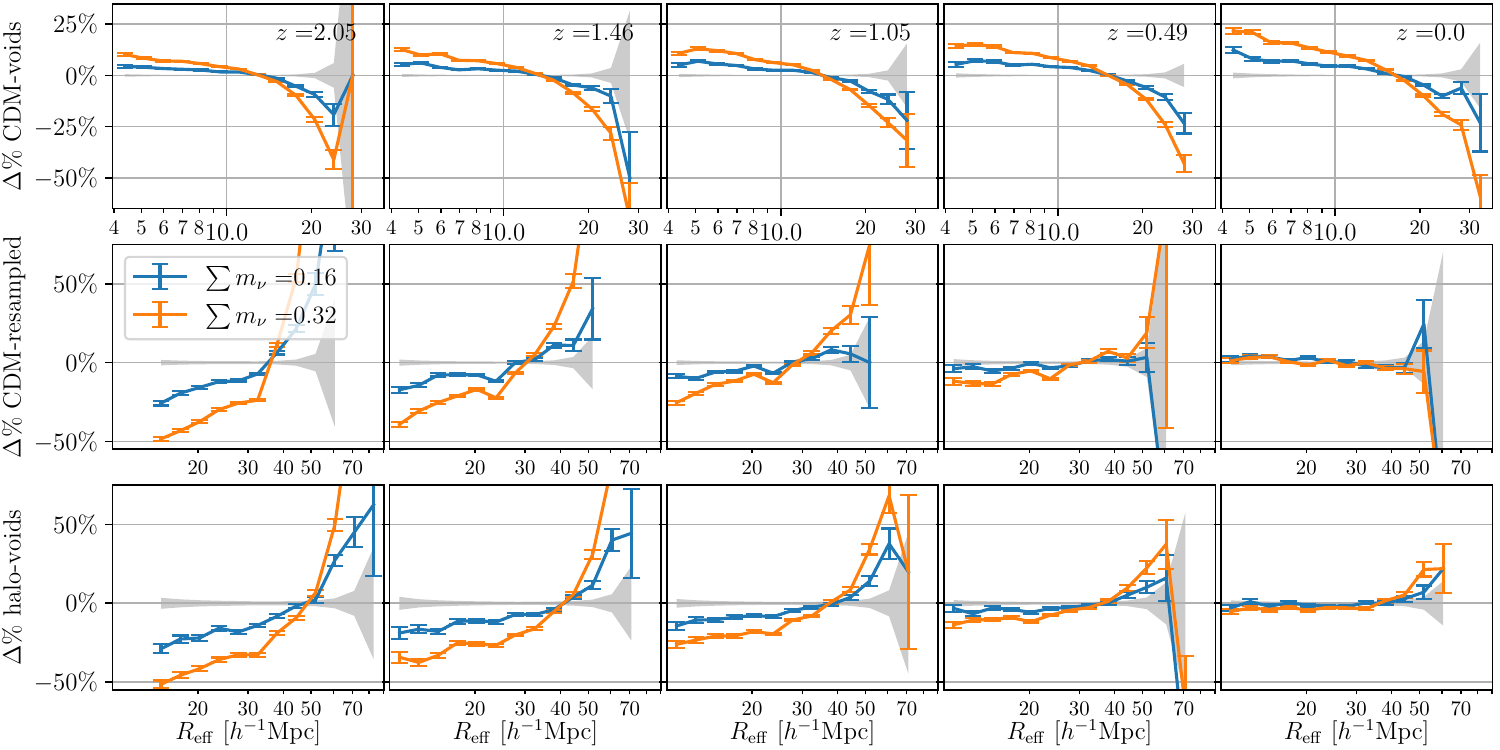}
\caption{VSF relative differences, with respect the \lcdm case, considering the cosmological constant case and neutrinos with total masses $\sum m_\nu=0.16$ (blue) and $0.32$ eV (orange), at $z=0$, 0.49, 1.05, 1.46. The upper panels show CDM-traced voids, the lower panels halo-traced voids for $M \geq 2.5 \times 10^{12} h^{-1} M_\odot$, the middle panels voids in the CDM distributions subsampled to match the corresponding halo density with mass. The errorbars are Poissonian errors, the grey shaded areas show the Poissonian errors for the \lcdm case. }
\label{fig:rel_neutrinos}
\end{figure}
\begin{figure}[t!]
\centering
\includegraphics[width=0.95\textwidth]{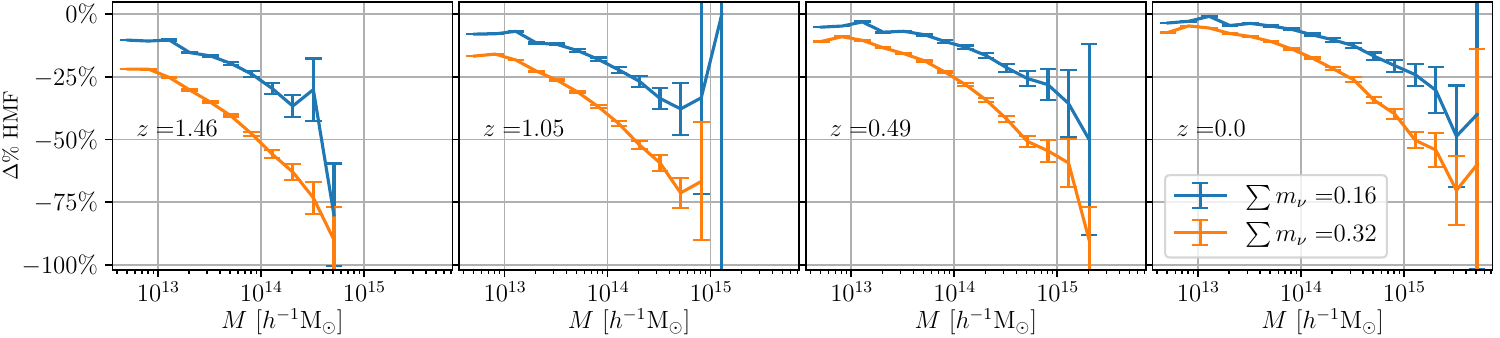}
\caption{Halo mass function relative differences, with respect the \lcdm case, considering the cosmological constant case and neutrinos with total masses $\sum m_\nu=0.16$ (blue) and $0.32$ eV (orange), at $z=0$, 0.49, 1.05, 1.46. The errorbars are Poissonian errors. }
\label{fig:HMF}
\end{figure}
We now explore the impact of the DE-EoS and the sum of neutrino masses, $\sum m_\nu$, on the VSF of watershed voids.
The effects on total neutrino mass on the evolution of the matter perturbations and on the background of the universe produce several effects. In particular, the total neutrino mass shows a degeneracy with respect to the amplitude of matter perturbations, $\sigma_8$, and to the matter density $\Omega_{\rm m}$. For this reason, in the literature there exist matched $\sigma_8(z=0)$ massless neutrino \lcdm simulation sets \citep[see e.g.][]{Quijote_2020}. This perspective has its own advantages, such as investigating the effect of massive neutrinos while mitigating the derived effect of the matter power spectrum normalisation. However, the decrease of $\sigma_8(z=0)$ in the presence of massive neutrinos is a physical effect of their free-streaming, so that $\sum m_\nu$ and $\sigma_8$ cannot be considered as independent parameters, especially if we notice that the amplitude of scalar perturbations $A_{\rm s}$ has been tightly constrained by CMB experiments \citep{planck2013,planck_2018}. If we ran simulations with the same total neutrino masses implemented in the DEMNUni set and $\sigma_8(z=0)$ normalised to the same massless-\lcdm value, $A_{\rm s}$ should be increased well beyond its constraints from CMB observations.  Furthermore, as discussed in \citep{parimbelli_2021}, Sec. 2.2, considering initial conditions  making $\sigma_8(z=0)$ to match the massless neutrino case would lead to cosmic history far from CMB and galaxy surveys observations. The initial conditions used for the DEMNUni runs have been chosen to match the Planck 2013 \citep{planck2013} $A_{\rm s}$ constraints, for this reason $A_{\rm s}$ is the same for all the DEMNUni cosmologies. This approach allows us to explore the global effects of massive neutrinos, both on the amplitude and shape of matter perturbations, given initial conditions physically motivated by CMB observations.

The investigation of DE-EoS and the sum of neutrino masses on the VSF was already explored in DEMNUni considering the impact of massive neutrinos~\citep{kreisch_2019} and dark energy~\citep{verza_2019}, separately. 
For the first time to date, we now extend these studies by considering the impact on the VSF of DE-EoS, massive neutrinos, and their combination. Furthermore, we investigate the effects of different tracers, the redshift dependence, and the geometrical distortions modifying the observed VSF. 

Fig.~\ref{fig:rel_neutrinos} shows the relative difference of the VSF measured in the presence of massive neutrinos, $\sum m_\nu=0.16\,{\rm eV}$ (blue) and $0.32\, {\rm eV}$ (orange), and in the cosmological constant case, i.e. $(w_0,w_{\rm a}) = (-1,0)$, with respect to \lcdm, i.e. massless neutrino case, at $z=0$, 0.49, 1.05, 1.46. The upper and lower panels show the results for CDM-voids and for halo-traced voids (with $M\geq 2.5 \times 10^{12} h^{-1} M_\odot$), while the middle panels show the results for voids in the CDM distributions subsampled to the corresponding halo number density, respectively.
For reference, it should be noted that the snapshot volume, i.e. 8 $(h^{-1}{\rm Gpc})^3$, roughly corresponds to the volume of a shell in the final data release of the Euclid survey centred at $z \sim 1$ and spanning $\pm \Delta z$ with $\Delta z = 0.1$~\citep{laureijs_20211_euclid_report}. 
The VSFs of halo- and CDM-traced voids are strongly impacted by massive neutrinos.\\
For CDM-traced voids, massive neutrinos suppress the number of large voids, shifting the VSF toward smaller radii with respect to the massless case. This is visible as the suppression of large voids and an increment of smaller voids. 
The change is due to the free-streaming of massive neutrinos: their thermal diffusion tends to smooth out matter density fluctuations in the Universe, suppressing their growth~\citep{lesgourgues_pastor_2006,lesgourgues_2013} and bringing matter in voids, which therefore become shallower and smaller with respect to the massless neutrino case \citep{massara_2015,kreisch_2019}.\\
For halo-traced voids the effect is the opposite:
massive neutrinos suppress halo formation too, therefore the population of haloes with the same mass has a different bias and mean tracer separation, both increasing with the total neutrino mass, with respect to the \lcdm case~\citep{Marulli_Carbone_2011}. To clarify this point, Fig.~\ref{fig:HMF} shows the relative difference of the halo mass function (HMF) measured in the presence of massive neutrinos with respect to the \lcdm case. It can be observed that haloes formation is suppressed by the presence of massive neutrinos, and therefore their number density decreases. As a consequence, the halo distribution with fixed minimum halo mass will trace larger scales as the neutrino mass increases, as extensively explained in the previous Section and showed in Fig.~\ref{fig:vide_voids_tracers_redshift_dep}. It follows that the size of halo-traced voids increases with the total neutrino mass. In order to isolate, from the tracer bias effect, the impact of the halo number density decrease induced by increasing neutrino masses, we consider the relative differences of the VSF in the CDM distributions subsampled to match the corresponding halo density (middle panels). Such differences show that the sparsity effect, due to neutrinos on tracer density above the masses considered in this work, inverts the relative VSF difference, being stronger than the direct impact of massive neutrinos on the CDM density fluctuations. The amplitude of the relative difference decreases with the redshift. This is due to the fact that the suppression of halo formation due to massive neutrinos lessens with decreasing redshift~\citep{Marulli_Carbone_2011}, as shown in Fig.~\ref{fig:HMF}. As expected, for halo-voids this effect dominates with respect to the one due to CDM density suppression affecting CDM-voids, and leads to the inversion of their relative VSFs~\citep{kreisch_2019}. On the other hand, it can be noticed that tracer bias slightly mitigates such inversion.

We note that massive neutrinos produce distinguishable effects on both the VSF of CDM- and halo-traced voids with respect to the massless case, for each neutrino mass and redshift considered. This result is particularly interesting since the minimum total neutrino mass allowed by neutrino flavour oscillations is roughly one-third of the minimum neutrino mass implemented in the DEMNUni simulations \citep{lesgourgues_pastor_2006,lesgourgues_2013,gerbino_lattanzi_2017}. Moreover, the minimum mass considered in the DEMNUni simulations is close to the upper limit constrained by Planck \cite{planck_2018}. This demonstrates the importance of using void statistics to analyse ongoing and upcoming galaxy surveys to constrain the sum of neutrino masses.
\begin{figure}[t!]
\centering
\includegraphics[width=0.95\textwidth]{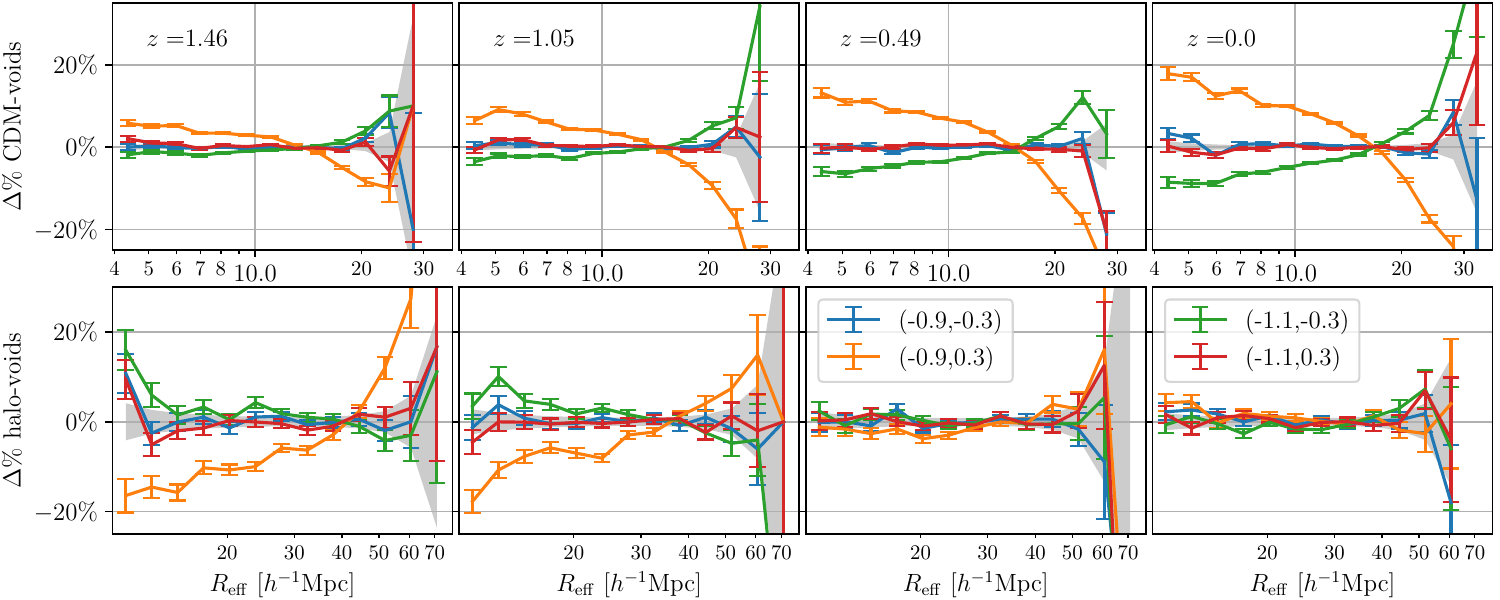}
\caption{VSF relative differences, with respect to the \lcdm case, for massless neutrinos and dynamical DE, at $z=0$, 0.49, 1.05, 1.46. The DE-EoS are $(w_0=-0.9,w_{\rm a}=-0.3)$ (blue), $(w_0=-0.9,w_{\rm a}=0.3)$ (orange), $(w_0=-1.1,w_{\rm a}=-0.3)$ (green), and $(w_0=-1.1,w_{\rm a}=0.3)$ (red). The upper panels show CDM-traced voids, the lower panels halo-traced voids for $M \geq 2.5 \times 10^{12} h^{-1} M_\odot$. The errorbars represent Poissonian errors, the grey shaded areas show the Poissonian uncertainty in the \lcdm case. 
}
\label{fig:rel_DE}
\end{figure}
Fig.~\ref{fig:rel_DE} shows the relative difference of the VSF in the presence of dynamical DE and massless neutrinos with respect to the \lcdm case, for $z=0$, 0.49, 1.05, 1.46. As before, the upper panels show the results for CDM-traced voids, the lower ones for halo-traced voids with $M\geq 2.5 \times 10^{12} h^{-1} M_\odot$. The impact of the considered DE-EoS on the VSF is globally smaller with respect to the effect of massive neutrinos, for both halo- and CDM-traced voids.\\
Concerning CDM-traced voids (upper panels), the impact of DE is well distinguishable for two of the four DE-EoS, corresponding to the two models less degenerate with the cosmological constant case, i.e. $(w_0=-0.9,w_{\rm a}=0.3)$ (orange lines) and $(w_0=-1.1,w_{\rm a}=-0.3)$ (green lines). 
The first of these two DE-EoS, i.e. $(w_0=-0.9,w_{\rm a}=0.3)$, is less negative than the cosmological constant EoS, i.e. $w(a)=-1$, at any epoch (see Fig.~\ref{fig:DE_EoS}, left panel). In the DEMNUni simulations the density parameters $\Omega_{\rm m}$ and $\Omega_{\rm DE}=1-\Omega_{\rm m}$ at $z=0$ are the same for all the cosmological models implemented, therefore this EoS results in an early-time dominating DE (see Fig.~\ref{fig:DE_EoS}, central panel). Since DE suppresses the growth of structures, an early-time dominating DE starts to slow down the growth of density fluctuations earlier than in \lcdm case (see Fig.~\ref{fig:DE_EoS}, right panel). 
The resulting VSF of CDM-traced voids is therefore shifted toward smaller radii than the \lcdm one. This effect is visible in the VSF ratios with respect to the \lcdm one (orange line), and is represented in the upper panels of Fig.~\ref{fig:rel_DE} as a decrease of large voids and an increase of smaller ones.  
The opposite is true for the $(w_0=-1.1,w_{\rm a}=-0.3)$ case (green line), representing an EoS that is always more negative than $-1$, corresponding to a late-time dominating DE. 
The separation between the VSF corresponding to different DE-EoS for CDM-traced voids increases as the redshift decreases. 
This is a consequence of the fact that the linear growth factors for the different DE-EoS get further apart from each other as time passes (right panel of Fig.~\ref{fig:DE_EoS}). \\
Concerning halo-traced voids (lower panels), the DE-EoS cases producing distinguishable effects on the VSF with respect to \lcdm are the same as for the CDM-traced, i.e. $(w_0=-0.9,w_{\rm a}=0.3)$ (orange lines) and $(w_0=-1.1,w_{\rm a}=-0.3)$ (green lines). Nevertheless, the effects are opposite with respect to CDM-traced voids. In the $(w_0=-0.9,w_{\rm a}=0.3)$ case, the number of large voids is enhanced and the number of smaller voids is suppressed with respect to the \lcdm case, the opposite happens for the $(w_0=-1.1,w_{\rm a}=-0.3)$ EoS. We note that also the redshift dependence shows an opposite trend with respect to CDM-traced voids: the difference between the VSF for the halo-traced voids, due to the different EoS, decreases with redshift and all the VSFs are indistinguishable at $z=0$.

Analogously to the massive neutrinos case, the inversion of trend in the VSF of halo-traced voids, with respect to the CDM-traced voids, is linked to the impact of DE on halo formation. An early-time dominating dark energy suppresses halo formation with respect to the \lcdm case, therefore, both the bias and the mean halo separation increase. This reflects in an increase of the natural scale at which the underlying fluctuations are traced by haloes.
The combination of the effects of DE on both the halo distribution and the underling matter fluctuations leads to this inversion of trend in the size function of halo-traced voids with respect to CDM-traced voids. More precisely, at low redshifts the impact on halo-traced voids due to the effect of DE on the halo population compensates the direct impact of DE on CDM-traced voids and even overcomes it at high redshifts.

\section{Geometrical and redshift space distortions}\label{sec:AP_RSD}

When analysing real data, geometrical and redshift-space distortions (RSDs) affect cosmological measurements. We now explore how they impact the VSF and its sensitivity to the DE-EoS and the total neutrino mass. In cosmological observations, the position of tracers is measured in spherical coordinates where the distance variable is the redshift. 
The conversion from redshift to comoving position depends on the cosmological model and the parameters assumed. A wrong cosmology reflects in geometrical distortions, known as \AP effect \citep{AP1979}. Moreover, the observed redshift is a combination of the cosmological redshift and the Doppler shift along the line of sight due to peculiar motions of tracers~\citep{kaiser_1987,hamilton_1998}; this effect impacts the inferred tracer position introducing the so called RSDs. These distortions are observational effects, therefore in the following we will consider halo-traced voids only, which are more close to voids detected in galaxy surveys.

We start by considering the effect of geometrical distortions on the measured VSF. In the previous Section, we investigated the impact of DE and massive neutrinos on the VSF. Nevertheless, the sensitivity of the VSF to the DE-EoS and total neutrino mass has to be considered as the combination of the DE and neutrino effects on the VSF (without accounting for geometrical and RSD distortions), explained above, with, additionally, the impact that a wrongly assumed DE-EoS has on void measurements and, therefore, on the observed VSF\footnote{Geometrical distortions due to massive neutrinos are negligible, as discussed later.}. 
More precisely, measuring the VSF to probe the DE-EoS also requires inferring the size of voids by assuming a cosmological model, in this case a specific DE-EoS. 
It follows that the sensitivity of the VSF (accounting for the impact of geometrical and RSD distortions) to the DE-EoS is a nonlinear function of the DE impact on the VSF itself (without accounting for geometrical and RSD distortions). In practice, neglecting for the moment RSD effects, the sensitivity of the VSF on the DE-EoS can be quantified by considering the \AP effect. This is obtained by assuming a reference cosmology in performing cosmological measurements and evaluating geometrical distortions due to a true cosmological model different from the assumed reference one. 
The impact of the distortion can be derived as \citep{ballinger_1996, eisenstein_2005_BAO, xu_padmanabhan_2013, sanchez_2017_BAO, hamaus_2020, correa_2021}
\begin{equation}\label{eq:AP}
r_\parallel' = \frac{H(z)}{H'(z)} \, r_\parallel = q^{-1}_\parallel \, r_\parallel 
\,, \quad r_\perp' = \frac{\chi'(z)}{\chi(z)} \, r_\perp = q^{-1}_\perp \, r_\perp \,;
\end{equation}
where $r_\parallel$ and $r_\perp$ are the comoving distances between two objects at redshift $z$ projected along the parallel and perpendicular directions with respect to the line-of-sight and separated by a small angle, $H(z)$ is the Hubble parameter, and $\chi(z)$ the comoving distance. The primed quantities refer to the ones computed in the reference cosmology, assumed when mapping redshifts to comoving distances, the non-primed ones correspond to the true cosmology. 
We note that for watershed voids, the redshift positions of all cells belonging to each basin do not vary under a smooth monotonic map. More importantly, the identification of density minima with the corresponding underdensity basins and Voronoi cells belonging to them is not affected, even if their shapes are distorted. 
This means that the redshift positions of the void centres extracted by \vide and the tracers belonging to each void are not affected by the assumed cosmology. 
Furthermore, the volume of each Voronoi cell estimated in a reference cosmology is modified according to $V_i' = q_\parallel^{-1} q_\perp^{-2} V_i$. It follows that the \AP effect can be computed exactly for \vide voids: the number of voids does not change when a wrong cosmology is used to infer distances, and the void effective radius $R_{\rm eff}$ appears modified according to $R'_{\rm eff} = q_\parallel^{-1/3}q_\perp^{-2/3} R_{\rm eff}$. The map from redshift to comoving distance impacts the inferred survey volume too~\citep{euclid_vsf}, and consequently the estimation of the number density of detected voids. The survey volume is $\Omega_{\rm [rad]} [\chi(z_{\rm out})^3 - \chi(z_{\rm in})^3]/3$, where $\Omega_{\rm [rad]}$ is the solid angle of the survey in steradians, while $z_{\rm in}$ and $z_{\rm out}$ are the redshift limits of the redshift range considered. Since here we work with comoving snapshots, we account for the change in volume effectively by considering the distortions along the Cartesian $z$-axis, so we implement the correction to the volume of the simulation box as
\begin{equation}
\frac{V_{\rm box}'}{V_{\rm box}} = \frac{\chi'^2(z)}{\chi^2(z)} \frac{H(z)}{H'(z)}\,.
\end{equation}

\begin{figure}[t!]
\centering
\includegraphics[width=0.95\textwidth]{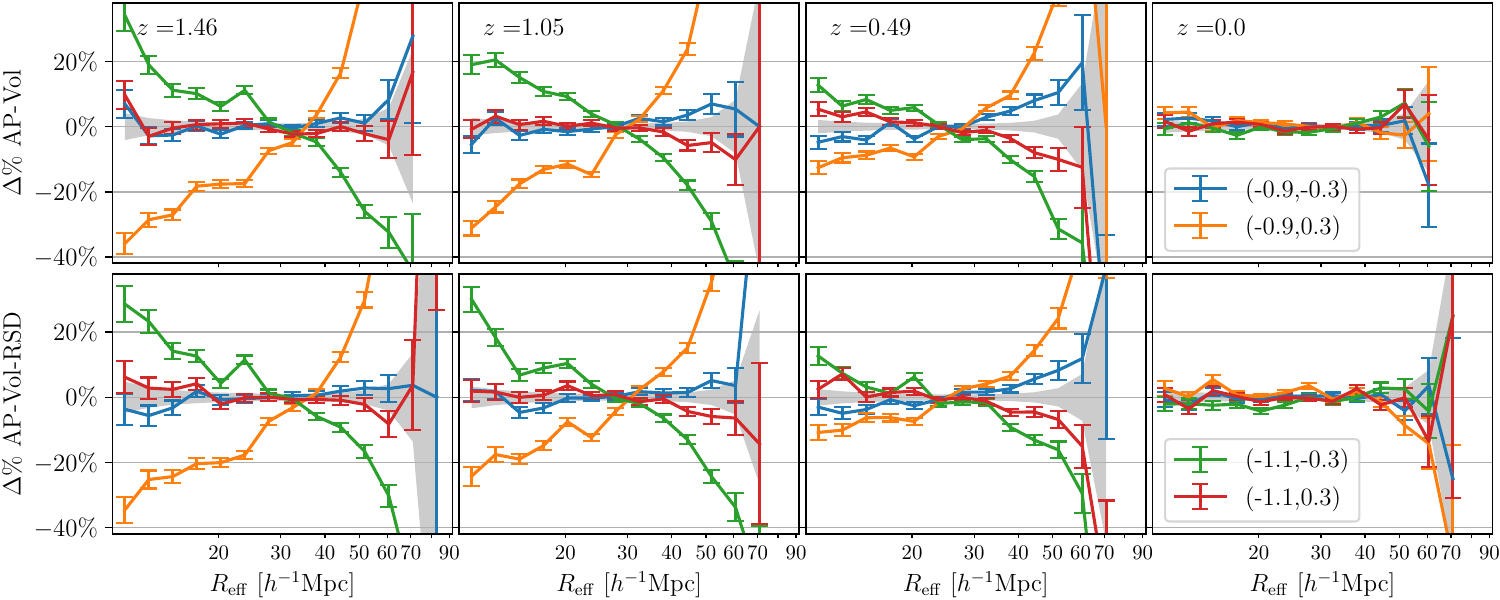}
\caption{Impact of geometrical distortions on the VSF relative differences, with respect to the \lcdm case, in dynamical DE simulations and massless neutrinos, at $z=0$, 0.49, 1.05, 1.46, for halo-traced voids detected in real space with geometrical distortions included, as discussed in the main text (upper panels) and redshift space (lower panels) with $M \geq 2.5 \times 10^{12} h^{-1} M_\odot$. The DE-EoS are: $(w_0=-0.9,w_{\rm a}=-0.3)$ (blue), $(w_0=-0.9,w_{\rm a}=0.3)$ (orange), $(w_0=-1.1,w_{\rm a}=-0.3)$ (green), and $(w_0=-1.1,w_{\rm a}=0.3)$ (red). The errorbars are Poissonian errors, the grey shaded areas show the Poissonian uncertainty for the \lcdm case.}
\label{fig:rel_DE_AP_RSD}
\end{figure}

The second source of distortions are RSDs, which impact the identification and observed properties of watershed voids. In particular, large voids in real space look larger in redshift space, due to the coherent streaming motion of tracers known as the Kaiser effect~\citep{kaiser_1987,hamilton_1998} which reflects in voids elongated along the line of sight~\citep{pisani_2015_RSD,hamaus_2015,hamaus_2020,correa_2021}, while the chaotic distribution of tracer peculiar velocities, known as Gaussian-streaming~\citep{fisher_1995_streaming,hamilton_1998,hamaus_2015,nadathur_2020,hamaus_2020,woodfinden_nadathur_2022} suppresses the number of smaller voids and makes less precise the identification of the void boundaries~\citep{pisani_2015_RSD}. RSDs depend on the cosmological model, at the linear level this dependence is fully encapsulated in the linear growth rate of density perturbations~\citep{hamilton_1998,hamaus_2015,cai_2016}. 
To account for the RSD impact on the VSF, we move the tracer positions along one Cartesian axis according to their peculiar velocity, and we then build a new \vide void catalogue. \\
Fig.~\ref{fig:rel_DE_AP_RSD} shows the impact of geometrical distortions (upper panels) plus RSD (lower panels) on the VSF relative differences, with respect to \lcdm case, in the presence of dynamical DE and massless neutrinos for halo-traced voids with $M\geq 2.5 \times 10^{12} h^{-1}M_\odot$ at $z=0$, 0.49, 1.05, 1.46. The assumed reference cosmology is the DEMNUni \lcdm one. 
The changes in the simulated real-space VSF for various DE-EoS induced by geometrical distortions (upper panels) break the degeneracy of all the considered DE-EoS. We note that at redshift $z=0$ there are no geometrical distortions, this is because the map between redshift and comoving distance converges to $\chi(z) \simeq cH_0 z$ and the entire DEMNUni set has $H_0= 67 \, {\rm km}\,{\rm s}^{-1}{\rm Mpc}^{-1}$.
The effect of geometrical distortions grows with redshift, enhancing the impact on the observed VSF where the wrong reference cosmology is assumed. The lower panels show the VSF relative difference with geometrical distortions combined with RSD. The effect on the VSF relative differences is to smear the separations between different cosmologies, nevertheless all the four EoS considered remain distinguishable in the VSF.

Geometrical distortions sourced by massive neutrinos can be neglected. As for the DE-EoS case, massive neutrinos impact the comoving distance estimation through the Hubble factor $H(z)$, in this way they can produce \AP distortion via Eq.~\eqref{eq:AP}. Nevertheless, this effect is negligible~\citep{zennaro_2017} for massive neutrinos at the redshift analysed in this work and, therefore, is not considered in the following. 

\begin{figure}[t!]
\centering
\includegraphics[width=0.95\textwidth]{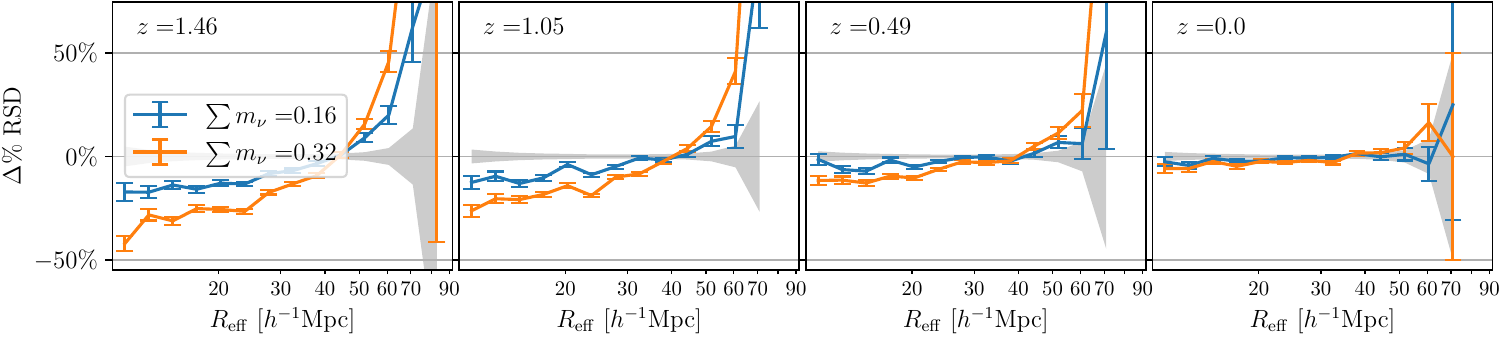}
\caption{Impact of RSDs on the VSF relative differences, with respect to the \lcdm case, in the presence of cosmological constant and massive neutrinos, $\sum m_\nu=0.16$ (blue) and 0.32 eV (orange), at $z=0$, 0.49, 1.05, 1.46, for halo-traced voids detected in redshift space with $M \geq 2.5 \times 10^{12} h^{-1} M_\odot$. The errorbars are Poissonian errors, the grey shaded areas show the Poissonian uncertainty for the \lcdm case.}
\label{fig:rel_MNU_RSD}
\end{figure}

Fig.~\ref{fig:rel_MNU_RSD} shows the VSF relative difference with respect to \lcdm for halo-traced voids with $M\geq 2.5 \times 10^{12} h^{-1}M_\odot$ measured in redshift-space, at $z=0$, 0.49, 1.05, 1.46. As in the real space VSF, shown in the lower panels of Fig.~\ref{fig:rel_neutrinos}, massive neutrinos produce distinguishable effects on the measured VSF, which are increased for large voids by RSDs.

\section{Decoupling dark energy from massive neutrinos}\label{sec:decoupling}

We now investigate the impact of the combination of DE and massive neutrinos on the VSF, and its capability to break the DE-massive neutrinos degeneracy when used to probe these components combined, also considering geometrical and RSD distortions. 

\begin{figure}[t!]
\centering
\includegraphics[width=0.95\textwidth]{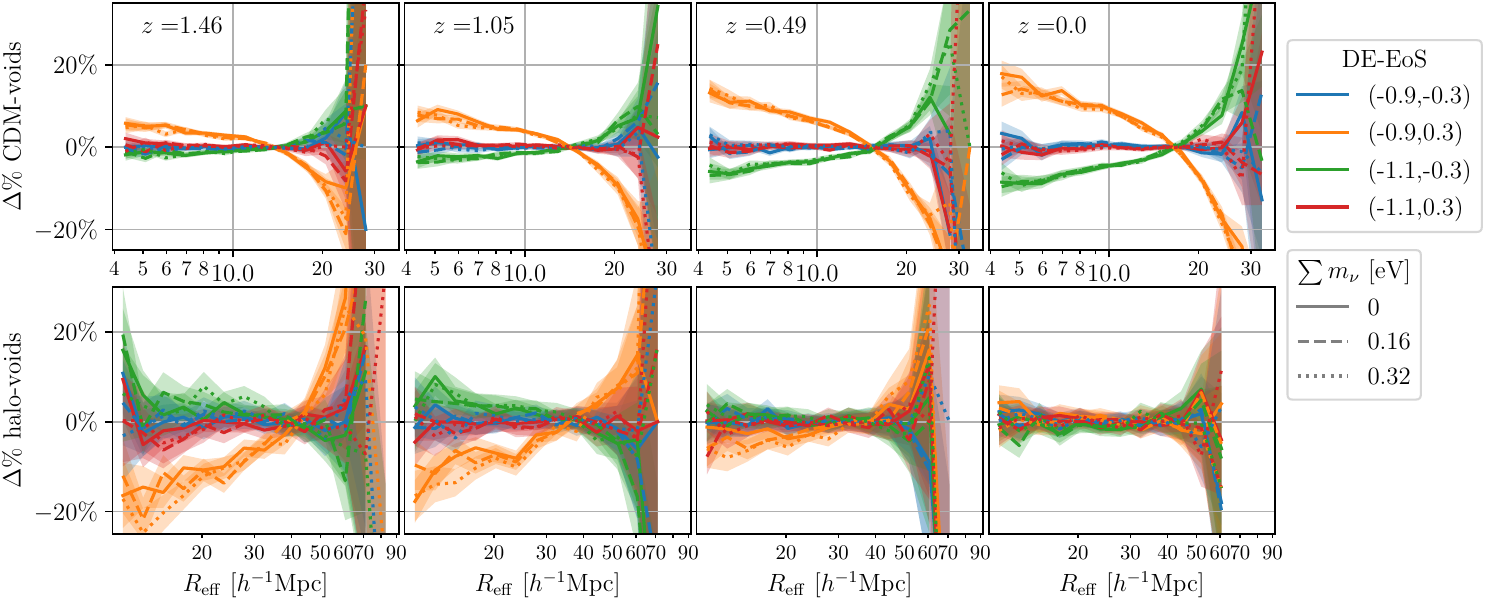}
\caption{For each fixed total neutrino mass, this plot shows the VSF relative difference in presence of dynamical DE with respect to the cosmological constant case at $z=0$, 0.49, 1.05, 1.46. Colour labels the DE-EoS as: $(w_0=-0.9,w_{\rm a}=-0.3)$ (blue), $(w_0=-0.9,w_{\rm a}=0.3)$ (orange), $(w_0=-1.1,w_{\rm a}=-0.3)$ (green), and $(w_0=-1.1,w_{\rm a}=0.3)$ (red); line style labels the total neutrino mass: 0 eV (solid), 0.16 eV (dashed), and 0.32 eV (dotted). The upper panels show CDM-traced voids, the lower panel halo-traced voids for $M \geq 2.5 \times 10^{12} h^{-1} M_\odot$. The errorbars are Poissonian errors. The overlap of the same colours and different line styles shows that DE-EoS and neutrino mass produce separable effects on the VSF.}
\label{fig:dec_DE_from_MNU}
\end{figure}
\begin{figure}[t!]
\centering
\includegraphics[width=0.95\textwidth]{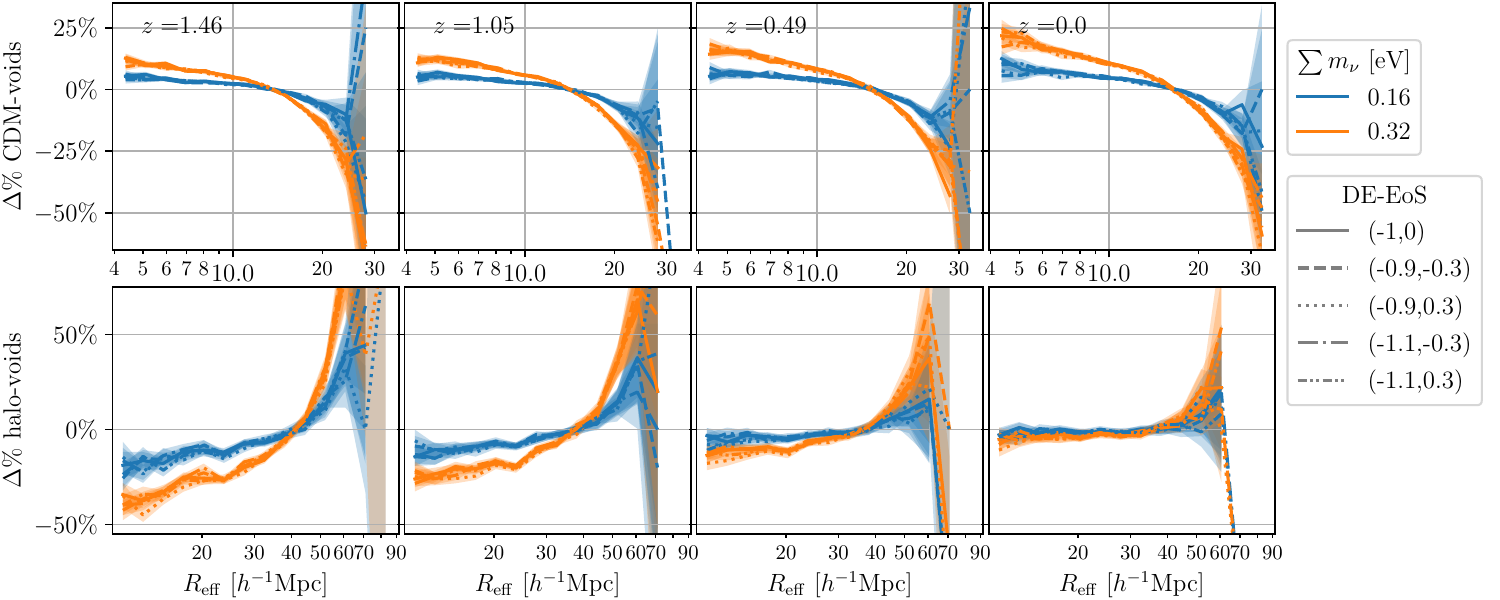}
\caption{For each fixed DE-EoS, this plot shows the VSF relative difference in the presence of massive neutrinos, $\sum m_\nu=0.16$ (blue) and 0.32 eV (orange), with respect the cosmological constant case at $z=0$, 0.49, 1.05, 1.46. Colours label the total neutrino mass: 0.16 eV (blue) and 0.32 eV (orange); Line style labels the DE-EoS: $(w_0=-1,w_{\rm a}=0)$ (solid), $(w_0=-0.9,w_{\rm a}=-0.3)$ (dashed), $(w_0=-0.9,w_{\rm a}=0.3)$ (dotted), $(w_0=-1.1,w_{\rm a}=-0.3)$ (dash-dotted), and $(w_0=-1.1,w_{\rm a}=0.3)$ (dash-dot-dotted). The upper panels show CDM-traced voids, the lower panels halo-traced voids for $M \geq 2.5 \times 10^{12} h^{-1} M_\odot$. The errorbars are Poissonian errors. The overlap of curves with the same colours and different line styles shows that DE-EoS and neutrino mass produce separable effects on the VSF.}
\label{fig:dec_MNU_from_DE}
\end{figure}

Let us first consider how the combination of DE-EoS and the total neutrino mass impacts the VSF when not accounting for geometrical and RSD distortions. To show that the effect of DE-EoS on the VSF is separable from the one of massive neutrinos, we proceed in the following way. First, we consider the impact of DE-EoS on the VSF when the total neutrino mass is fixed. We fix the total neutrino mass and we measure the relative values of the VSF corresponding to various DE-EoS with respect to the cosmological constant case. This is done for all the three neutrino masses explored in DEMNUni simulations. 
Each panel of Fig.~\ref{fig:dec_DE_from_MNU} disentangles the effect of the DE-EoS with respect to the sum of neutrino masses at a different redshift for both the CDM (top) and halo (bottom) voids. For each redshift we show the VSF relative difference with respect to the cosmological constant case when fixing the neutrino mass but changing the DE-EoS. 
Different line styles correspond to considering a different fixed sum of neutrino masses. Since different line styles are compatible within errorbars, different DE-EoS can be distinguished independently from what the considered value of the sum of neutrino masses is. In other words, the contribution of DE-EoS to the VSF is independent of the total neutrino mass.

We now consider the opposite case: we fix the DE-EoS and consider the impact of the neutrino mass only on the VSF. Analogously to the previous case for each fixed DE-EoS we measure the VSF relative difference in the presence of massive neutrinos with respect to the massless case. We then compare the VSF relative differences obtained in this way among the five DE-EoS explored in DEMNUni simulations. 
Each panel of Fig.~\ref{fig:dec_MNU_from_DE} disentangles the effect of the neutrino mass with respect to the DE-EoS at a different redshift for both the CDM (top) and halo (bottom) voids. For each redshift we show the VSF relative difference with respect to the massless case when fixing the DE-EoS. Different line styles correspond to considering a different fixed DE-EoS. Since different line styles are compatible within errorbars, the sum of neutrino masses can be distinguished independently from what the DE-EoS is. This result further confirms that DE-EoS and massive neutrinos produce separable effects on the VSF, for both CDM- and halo-traced voids.

\begin{figure}[t!]
\centering
\includegraphics[width=0.95\textwidth]{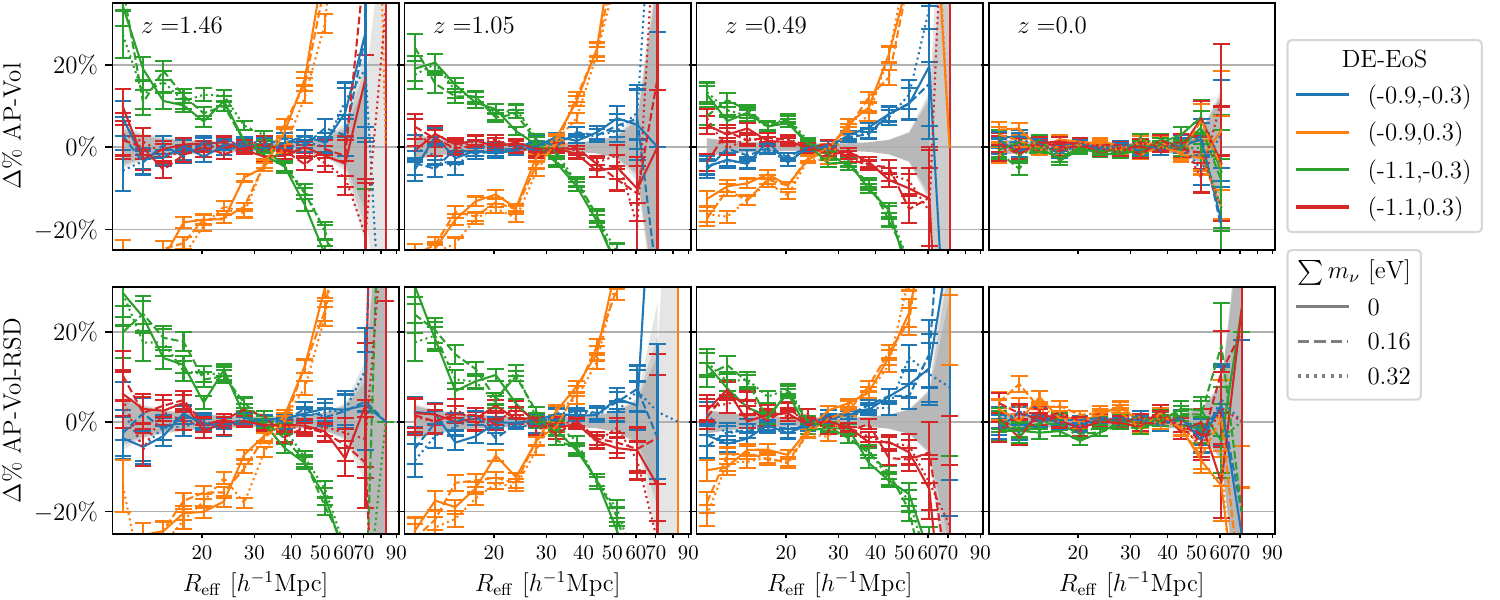}
\caption{For each fixed total neutrino mass, this plot shows the VSF relative difference for halo-traced voids with $M \geq 2.5 \times 10^{12} h^{-1} M_\odot$ in presence of dynamical DE with respect to the cosmological constant case considering geometrical effects (upper plots) plus RSDs (lower plots) at $z=0$, 0.49, 1.05, 1.46. Colour labels the DE-EoS as: 
$(w_0=-0.9,w_{\rm a}=-0.3)$ (blue), $(w_0=-0.9,w_{\rm a}=0.3)$ (orange), $(w_0=-1.1,w_{\rm a}=-0.3)$ (green), and $(w_0=-1.1,w_{\rm a}=0.3)$ (red); line style labels the total neutrino mass: 0 eV (solid), 0.16 eV (dashed), and 0.32 eV (dotted). The errorbars are Poissonian errors. The overlap of the same colours and different line styles shows that DE-EoS and neutrino mass produce separable effects on the VSF.}
\label{fig:dec_DE_from_MNU_AP_RSD}
\end{figure}

To explore the capability of the VSF to probe both DE and massive neutrinos and to break the degeneracy of their combined effects, we consider how the VSF is impacted by geometrical and redshift-space distortions, focusing on halo-traced voids only. As discussed above, massive neutrinos do not source geometrical distortions. 
For each fixed total neutrino mass, Fig.~\ref{fig:dec_DE_from_MNU_AP_RSD} shows the impact of geometrical distortions (upper panels) plus RSDs (lower panels) on the VSF relative difference in the presence of DE with respect to the cosmological constant case, for halo-traced voids with $M\geq 2.5 \times 10^{12} h^{-1}M_\odot$. The corresponding DE-EoS is represented by the colour and the total neutrino mass by the line style, listed in the legend. We note that the global effect of DE on the VSF can be separated from the one of massive neutrinos. This is expected since in the VSF, even without accounting for geometrical and RSD distortions, DE-EoS effects can be separated from massive neutrinos effects, and DE only induces geometrical distortions. 
We note, however, that geometrical distortions strongly increase the sensitivity to DE-EoS, allowing to distinguish among all the considered DE-EoS when the total neutrino mass is fixed.
We note that since DE-EoS and massive neutrinos effects are separable on the measured VSF, and massive neutrinos do not source geometrical distortion, it follows that the global effect on the VSF is the sum of the DE one explored here plus the one of massive neutrinos, shown in the lower panels of Fig.~\ref{fig:rel_neutrinos}. Therefore the VSF can be used to distinguish among all the combinations of DE-EoS and total neutrino masses considered. 
This result is a consequence of the capability of the VSF to distinguish among the considered DE-EoS and of the separability of DE from massive neutrinos effects on the VSF, together with the fact that DE and massive neutrinos effects on the VSF show different redshift dependencies. 
The lower panels show the impact of geometrical distortions together with RSD. We note that the relative difference is reduced with respect to the real-space case, but all the combinations of DE-EoS and total neutrino mass considered remain distinguishable.

\begin{figure}[t!]
\centering
\includegraphics[width=0.95\textwidth]{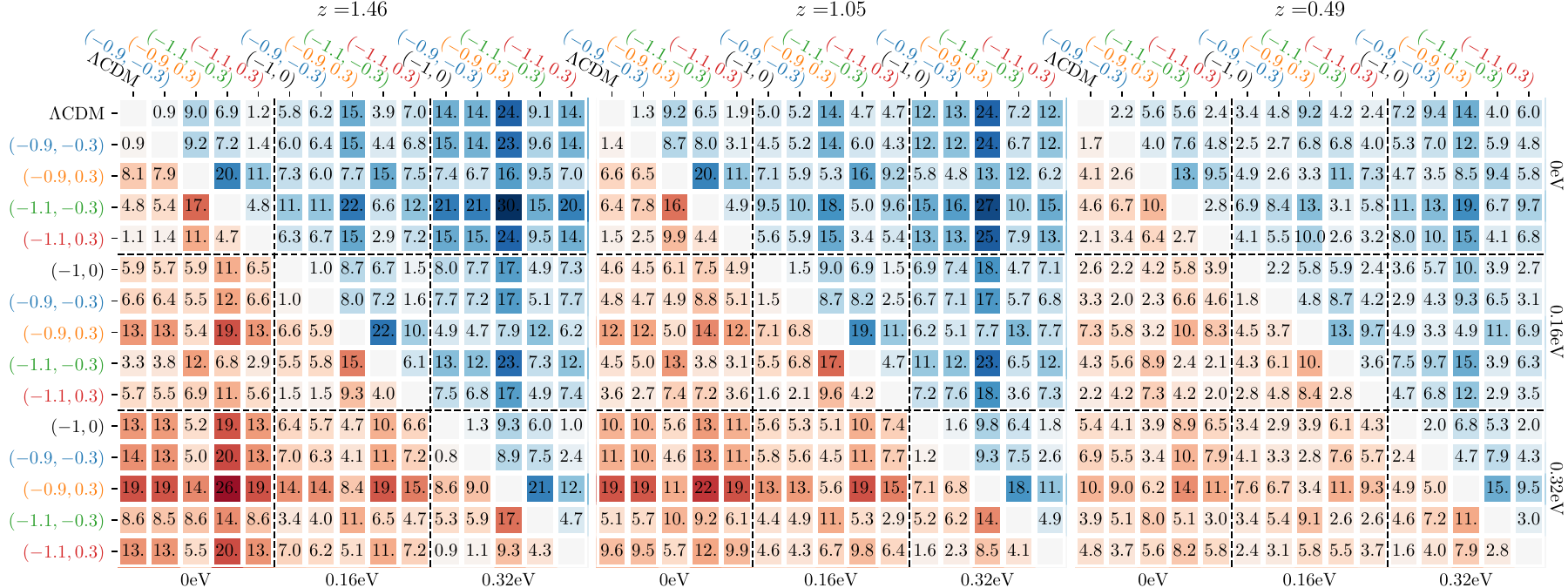}
\caption{${\cal R}_{i\text{-}j}$ value for the VSF relative difference corresponding to each pair of all the possible combinations of DE-EoS and total neutrino mass considered, for halo-traced voids with $M \geq 2.5 \times 10^{12} h^{-1} M_\odot$, considering geometrical (blue upper triangular matrices) plus RSD (red lower triangular matrices) distortions at $z=0.49$, 1.05, 1.46. Each row and column are labelled with the corresponding DE-EoS, with the exception of \lcdm for clarity; the text colour corresponds to the one used to represent the different DE-EoS in the other plots. The total neutrino mass corresponding to each DE-EoS is specified on the opposite matrix side with respect to the DE-EoS labels.}
\label{fig:comparison_all_table}
\end{figure}

To better visualise and quantify the capability of the VSF to distinguish among all combinations of DE-EoS and total neutrino mass considered, for each pair of the combinations considered, we compute the quantity
\begin{equation}
{\cal R}_{i\text{-}j} = \frac{1}{N_{\rm bins}}\sum_{k=0}^{N_{\rm bins}} \frac{1}{\sigma_k^{(i\text{-}j)}}\left\vert\frac{{\rm VSF}_{i,k}}{{\rm VSF}_{j,k}}-1\right\vert \,,
\end{equation}
where ${\rm VSF}_{i,k}$ and ${\rm VSF}_{j,k}$ are the VSF measured in the $k^{\rm th}$ radius bin for the two combinations of DE-EoS and total neutrino masses we are comparing, labelled $i$ and $j$; $N_{\rm bins}$ is the number of radius bins considered; $\sigma_k^{(i\text{-}j)}$ is the relative error of the two VSFs considered in the $i^{\rm th}$ radius bin. The quantity ${\cal R}_{i\text{-}j}$ allows us to quantify whether the VSF corresponding to two combinations of DE-EoS and total neutrino mass can be distinguished. This is formally the signal-to-noise ratio of the measured VSF to distinguish among different cosmological parameters: ${\cal R}_{i\text{-}j}<1$ means that the two cosmologies are not distinguishable using the VSF; when ${\cal R}_{i\text{-}j}>1$, the higher the ${\cal R}_{i\text{-}j}$ is, the better they can be distinguished with the VSF. Fig.~\ref{fig:comparison_all_table} shows the ${\cal R}_{i\text{-}j}$ value for each possible combination of DE-EoS and total neutrino mass, for halo-traced voids with $M>2.5 \times 10^{12} h^{-1}M_\odot$, at $z=0.49$, 1.05, 1.46. The blue upper triangular matrices show the results in real-space considering the impact of geometrical distortions, the red triangular matrices show the results in redshift-space considering geometrical distortions as well. We note that all the pairs of VSFs corresponding to each possible combination of DE-EoS and total neutrino mass are distinguishable both in real- and in redshift-space. If a pair is indistinguishable at a given redshift, i.e. ${\cal R}_{i\text{-}j}<1$, they disentangle at a different redshift.

Moreover, the ${\cal R}_{i\text{-}j}$ quantity showed in Fig.~\ref{fig:comparison_all_table} allows us to quantify the relative difference between the VSF shown and discussed above, i.e. in Secs.~\ref{sec:DE_MNU_VSF},~\ref{sec:AP_RSD}, and this one. In particular, the first row (real space) and the first column (redshift space) of the first block, i.e. the comparison of the four DE-EoS with $\sum m_\nu=0$ with respect to \lcdm, correspond to the upper and lower panels of Fig.~\ref{fig:rel_DE_AP_RSD}, respectively. Focusing on real-space, i.e. the blue upper triangular matrices, the first element of each total neutrino mass block along the first row quantifies what is shown in the lower panels of Fig.~\ref{fig:rel_neutrinos}, i.e. the comparison between cosmologies with $\sum m_\nu>0$ and $(w_0,w_{\rm a})=(-1,0)$ and \lcdm. Concerning the corresponding redshift-space counterpart, i.e. the red lower triangular matrices, the first element of each block along the first column refers to Fig.~\ref{fig:rel_MNU_RSD}. These matrices also quantify the separability of DE-EoS and massive neutrino effects in the presence of geometrical and redshift-space distortions, discussed in this Section: the diagonal of each neutrino mass block shows the VSF relative difference in the presence of fixed DE-EoS with respect to different neutrino mass. It can be noted that for each block the values along the block diagonal are almost the same, as a confirmation of the separability of the DE and massive neutrinos effects on the VSF discussed before. In particular, the values in the real-space (blue) part correspond to the lower panels of Fig.~\ref{fig:dec_MNU_from_DE}. The first row (column) of each neutrino mass block along the diagonal quantifies the separability of the VSFs corresponding to each DE-EoS with respect to the cosmological constant for fixed neutrino mass in the presence of geometrical distortions (plus RSDs). These results correspond to the upper (lower) panels of Fig.~\ref{fig:dec_DE_from_MNU_AP_RSD}, we note that also in this case the values of different blocks are almost the same.

\section{Conclusions}

In this work, we have shown that the VSF is a sensitive probe to constrain the DE-EoS and to measure the sum of neutrino masses. In particular, our analysis has shown that the VSF is able to distinguish among each combination of total neutrino mass and DE-EoS considered in this work. 

We extended previous results concerning DE \citep{verza_2019} and massive neutrinos \citep{kreisch_2019} by considering various redshifts and tracers, evaluating the impact of geometrical and redshift-space distortions on the measured VSF, and giving insights into the physical processes affecting the corresponding VSF. In particular, we found that RSDs partially reduce the relative difference sourced by the DE-EoS and the total neutrino mass in the VSF, nevertheless without cancelling out their differences. On the other hand, geometrical distortions enhance the effects of DE on the VSF, allowing us to break the existing degeneracies between different DE-EoS.

In addition, we have shown that, considering both geometrical and RSD effects, the VSF can distinguish among all the combinations of DE-EoS and total neutrino masses explored in this work.
This is the main result of our analysis, which shows that the VSF can break existing degeneracies between different DE-EoS and total neutrino masses. 

We point out here that the choice of a fixed $A_s$ may correlate void statistics with CMB anisotropies. Therefore, in the Appendix we perform the analysis by imposing that all the cosmologies share the same $\sigma_8$ at $z=0$, obtained as a derived parameter from Planck data. While the latter choice may implicitly impose a further prior on the cosmological model via which the $\sigma_8$ value is derived, we show in the Appendix that such a choice does not impact the results presented in this work, rather, in some cases, the sensitivity of the VSF to cosmological parameters is even increased. In addition, we show how the uncertainty on $A_s$ (and therefore on the derived $\sigma_8$ value) from Planck constraints, propagates to errors on the VSF and compare those with Poissonian errors from the measurements presented in this work.

In this context, we note that the volume and resolution of DEMNUni simulations mimic those of upcoming galaxy surveys, and the minimum total neutrino mass implemented is of the same magnitude as upper bounds measured by the Plank Collaboration \citep{planck_2018}, being $\sim 2.5$ the minimum total neutrino mass allowed by neutrino oscillations. Moreover, the four DE-EoS explored beyond the cosmological constant case are within the current constraints \citep{planck_2018} and challenging to be detected even with future galaxy survey analyses~\citep{euclid_IST_fishers_2020}. This demonstrates the importance of the VSF as a probe in galaxy surveys, possibly in combination with other cosmological probes, to measure the total neutrino mass and constrain the DE-EoS. 

This work, exploiting measurements of large simulations, shows that the VSF is a promising tool for cosmological analysis. On the other hand, the available theoretical VSF model \citep{SVdW,jennings2013}, when used for data analysis, requires to dramatically reduce the available void statistics and consequently its power in constraining cosmology \citep[see e.g. discussion in][]{verza_2019,euclid_vsf}.
A way to extend this work is to consider other possibilities to model the VSF, designed to maximise the use of available void statistics.

A further application could be the combination of the VSF with other (void) statistics, such as the void-galaxy cross-correlation. This quantity is used to probe the cosmological model via RSD and the \AP effect~\citep{hamaus_2020,nadathur_2020,hamaus_2022_euclid}. The VSF in combination with the void-galaxy cross-correlation can provide a tool to constrain the DE-EoS and the total neutrino mass, as well as other cosmological parameters and the expansion history of the Universe, possibly breaking some of the existing degeneracies~\citep{bayer_2021,kreisch_2021,pellicciari_2022}.

Finally, the results presented in this work show the 0-order statistics of voids, i.e. the VSF. A possible way to extend this work is to explore higher statistics of voids, such as the void-void and void-tracer correlation functions, and their sensitivity to distinguish among various combinations of DE-EoS and total neutrino masses.

\acknowledgments 
GV and AR are supported by the project ``Combining Cosmic Microwave Background and Large Scale Structure data: an Integrated Approach for Addressing Fundamental Questions in Cosmology'', funded by the MIUR Progetti di Rilevante Interesse Nazionale (PRIN) Bando 2017 - grant 2017YJYZAH.
AR acknowledges funding from Italian Ministry of Education, University and Research (MIUR) through the `Dipartimenti di eccellenza' project Science of the Universe. 
AP is supported by NASA ROSES grant 12-EUCLID12-0004, and NASA grant 15-WFIRST15-0008 to the Nancy Grace Roman Space Telescope Science Investigation Team ``cosmology with the High Latitude Survey''. AP acknowledges support from the Simons Foundation to the Center for Computational Astrophysics at the Flatiron Institute.
The DEMNUni simulations were carried out in the framework of ``The Dark Energy and Massive-Neutrino Universe" project, using the Tier-0 IBM BG/Q Fermi machine and the Tier-0 Intel OmniPath Cluster Marconi-A1 of the Centro Interuniversitario del Nord-Est per il Calcolo Elettronico (CINECA). We acknowledge a generous CPU and storage allocation by the Italian Super-Computing Resource Allocation (ISCRA) as well as from the coordination of the ``Accordo Quadro MoU per lo svolgimento di attività congiunta di ricerca Nuove frontiere in Astrofisica: HPC e Data Exploration di nuova generazione'', together with storage from INFN-CNAF and INAF-IA2.

\bibliographystyle{JHEP}
\bibliography{biblio}

\appendix

\section{VSF behaviour when normalising at $\sigma_8$}

In this appendix, we explore the effect of the $\sigma_8$ normalisation on the VSF by using the theoretical model of Sheth and Van de Weygaert~\citep{SVdW}. According to such a model, the VSF depends on the cosmology through the linear power spectrum, more precisely through $\sigma(R_{\rm L})$, i.e. the square root of the integral of the linear power spectrum convolved with a window function, 
\begin{equation}
\sigma^2(R_{\rm L}) = \int \frac{{\rm d} k}{2 \pi^2} k^2 P(k) |W(kR_{\rm L})|^2,
\end{equation}
where $R_{\rm L}$ is the Lagrangian radius. 
We have considered both the cases of varying the DE-EoS and $\sum m_\nu$.

\subsection{DE-EoS case}\label{sec:sig8_DE}
The DE-EoS affects the linear growth factor alone, i.e. it drives the redshift evolution of the global amplitude of the power spectrum, and therefore of $\sigma(R_{\rm L})$, while the power spectrum and $\sigma(R_{\rm L})$ shapes are not modified. 
Fig.~\ref{fig:Dz_normAsSig8} shows, for the four DEMNUni DE-EoS cosmologies, the effect of normalising the power spectrum to the same $\sigma_8$ (right panel) with respect to normalising it at the same amplitude, $A_s$, of the primordial scalar perturbations as measured by CMB experiments (left panel).
\begin{figure}[t!]
\centering
\includegraphics[width=0.95\textwidth]{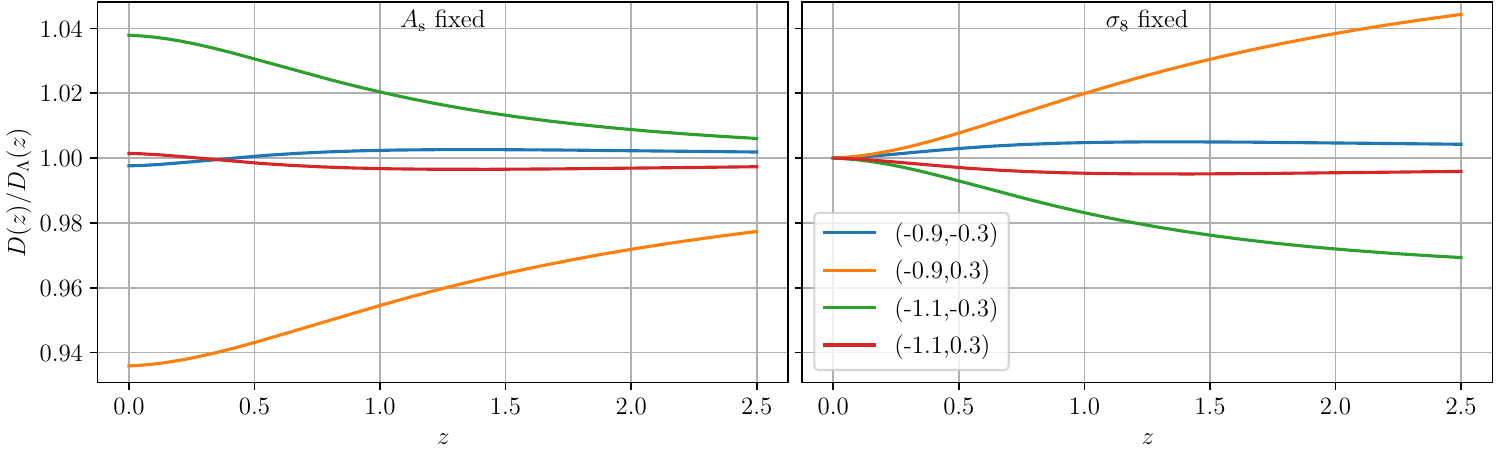}
\caption{Linear growth factors in the massless neutrino case for the considered DE-EoS with respect to \lcdm. The left panel shows when they are normalised at the amplitude of the primordial scalar perturbations ($A_{\rm s}$ fixed), the right panel when normalised at $z=0$ ($\sigma_8$ fixed). The colour labels the DE-EoS as: 
$(w_0=-0.9,w_{\rm a}=-0.3)$ (blue), $(w_0=-0.9,w_{\rm a}=0.3)$ (orange), $(w_0=-1.1,w_{\rm a}=-0.3)$ (green), and $(w_0=-1.1,w_{\rm a}=0.3)$ (red).}
\label{fig:Dz_normAsSig8}
\end{figure}
To explore the effect of fixing $\sigma_8$ instead of $A_{\rm s}$ we use the theoretical Sheth and Van de Weygaert~\citep{SVdW} model below: 
\begin{equation}\label{eq:SVdW}
\frac{{\rm d} n(R)}{{\rm d} R}  = \left. \frac{3}{4 \pi R_{\rm L}^3} f[\sigma(R_{\rm L})] \, \frac{{\rm d} \sigma(R_{\rm L})}{{\rm d} R_{\rm L}} \right\vert_{R_{\rm L}=R_{\rm L}(R)} \,,
\end{equation}
where
\begin{equation}
f(\sigma) = \frac{2}{\sigma} \sum_{j=1}^{\infty} \, \exp{\left[-\frac{(j \pi x)^2}{2}\right]} \, j \pi x^2 \, \sin{\left( j \pi \mathcal{D} \right)}\,, \qquad 
\mathcal{D} = \frac{|\delta_\mathrm{v}|}{\delta_\mathrm{c} + |\delta_\mathrm{v}|}\, , \qquad x = \frac{\mathcal{D}}{|\delta_\mathrm{v}|} \sigma \, .
\end{equation}
In this model $\delta_{\rm c}$ is the linear threshold for halo collapse, $\delta_{\rm v}$ is the linear void formation threshold, $R_{\rm L}$ is the Lagrangian void radius, $R$ is the evolved one that can be derived as $R = (1 + \delta_{\rm v}^{\rm NL})^{-1/3} R_{\rm L}$, where $\delta_{\rm v}^{\rm NL}$ is the nonlinear density contrast corresponding to $\delta_{\rm v}$. 
We consider $\delta_{\rm v}$ a free parameter that we fit at each $z$ to the void size function measured in the \lcdm simulation using the corresponding $\sigma(R_{\rm L},z)$, computed as $\sigma(R_{\rm L},z) = \sigma(R_{\rm L},z=0)  D(z)$, being $D(z)$ the linear growth factor.
As a second step, we use the $\sigma(R_{\rm L},z)$ corresponding to the DE-EoS investigated in this work, verifying that the behaviour, amplitude, and uncertainties of the void size function residuals, reported in Fig.~\ref{fig:rel_DE}, are well reproduced. Finally, we use the \lcdm matter power spectrum at $z=0$ to compute $\sigma(R_{\rm L},z=0)$, and then evolve the latter using the $D(z)$ corresponding to the different DE-EoS in order to obtain the $\sigma_8$-matched $\sigma(R_{\rm L},z)$. We then use this quantity to compute the corresponding VSF in the DE case.

We repeated the same procedure both for CDM- and halos-voids, in the latter case also considering the halo bias effect accounted as follows: i) we fit the threshold, $\delta_{\rm v}$, to the \lcdm simulated data; ii) using the methodology described in \cite{verza_2019}, we compute the nonlinear counterpart, $\delta_{\rm v}^{\rm NL}$; iii) we find the corresponding density contrast in the halo distribution multiplying $\delta_{\rm v}^{\rm NL}$ by the analytical effective halo bias from \cite{sheth_tormen_2001}; iv) we derive the corresponding nonlinear matter density contrast for the different DE-EoS as $\delta_{\rm v}^{\rm NL} b_{\rm eff}^{\rm \Lambda CDM} / b_{\rm eff}^{w_0,w_{\rm a}}$; v) finally, using the methodology in \cite{verza_2019} in the backword direction, we obtain the corresponding linear quantity and insert it in Eq.~\eqref{eq:SVdW}.
In Fig.~\ref{fig:rel_DE_sig8} we show the results: as expected, the behaviour of the relative differences is inverted with respect to the cases studied in this work, since the different normalisation, with respect to $\sigma_8$ rather than $A_s$, inverts the $D(z)$ trend with respect to the \lcdm case (see Fig.~\ref{fig:Dz_normAsSig8}). 
Moreover, Fig.~\ref{fig:rel_DE_sig8} shows that, by fixing $\sigma_8$, the difference between the two most degenerate cases increases even more. This can be explained by considering the growth factor behaviour in Fig.~\ref{fig:Dz_normAsSig8}.

It follows that the blue and red cases, i.e. $(w_0,w_{\rm a})=(-0.9,-0.3)$ and $(-1.1,0.3)$, are equal at $z=0$. This means that they would not cross each other during the redshift evolution, maximising the difference between themselves and the \lcdm case as the redshift increases. The other two cases roughly maintain a difference of the same order of magnitude as in the case of fixing the normalisation to $A_s$, but with an inverted redshift dependence.
\begin{figure}[t!]
\centering
\includegraphics[width=0.95\textwidth]{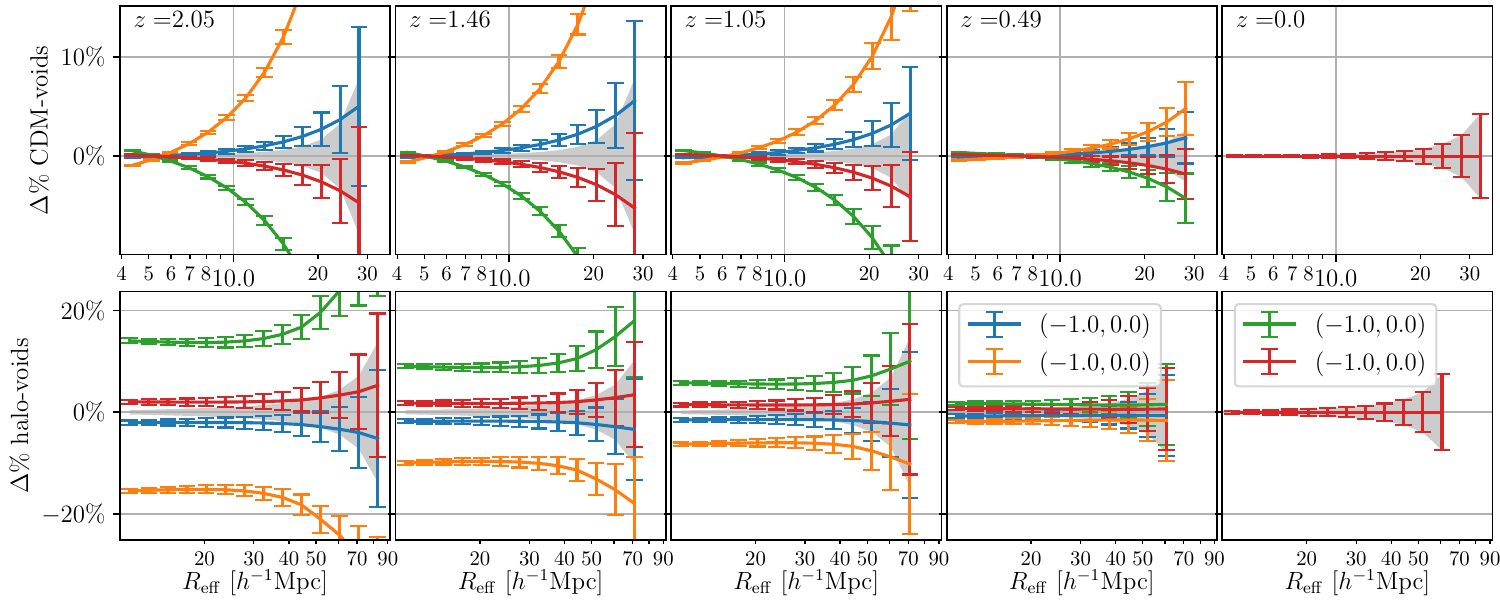}
\caption{VSF relative differences, with respect to the \lcdm case, for massless neutrinos and dynamical DE, at $z=0$, 0.49, 1.05, 1.46. For each DE-EoS the VSF is computed with the Sheth and Van de Waygaert model~\citep{SVdW} with the power
spectrum normalised at $\sigma_8=0.83$, as discussed in the text. The DE-EoS are $(w_0=-0.9,w_{\rm a}=-0.3)$ (blue), $(w_0=-0.9,w_{\rm a}=0.3)$ (orange), $(w_0=-1.1,w_{\rm a}=-0.3)$ (green), and $(w_0=-1.1,w_{\rm a}=0.3)$ (red). The upper panels show CDM-traced voids, the lower panels halo-traced voids for $M \geq 2.5 \times 10^{12} h^{-1} M_\odot$. The errorbars represent Poissonian errors, the grey shaded areas show the Poissonian uncertainty in the \lcdm case. 
}
\label{fig:rel_DE_sig8}
\end{figure}
\begin{figure}[t!]
\centering
\includegraphics[width=0.95\textwidth]{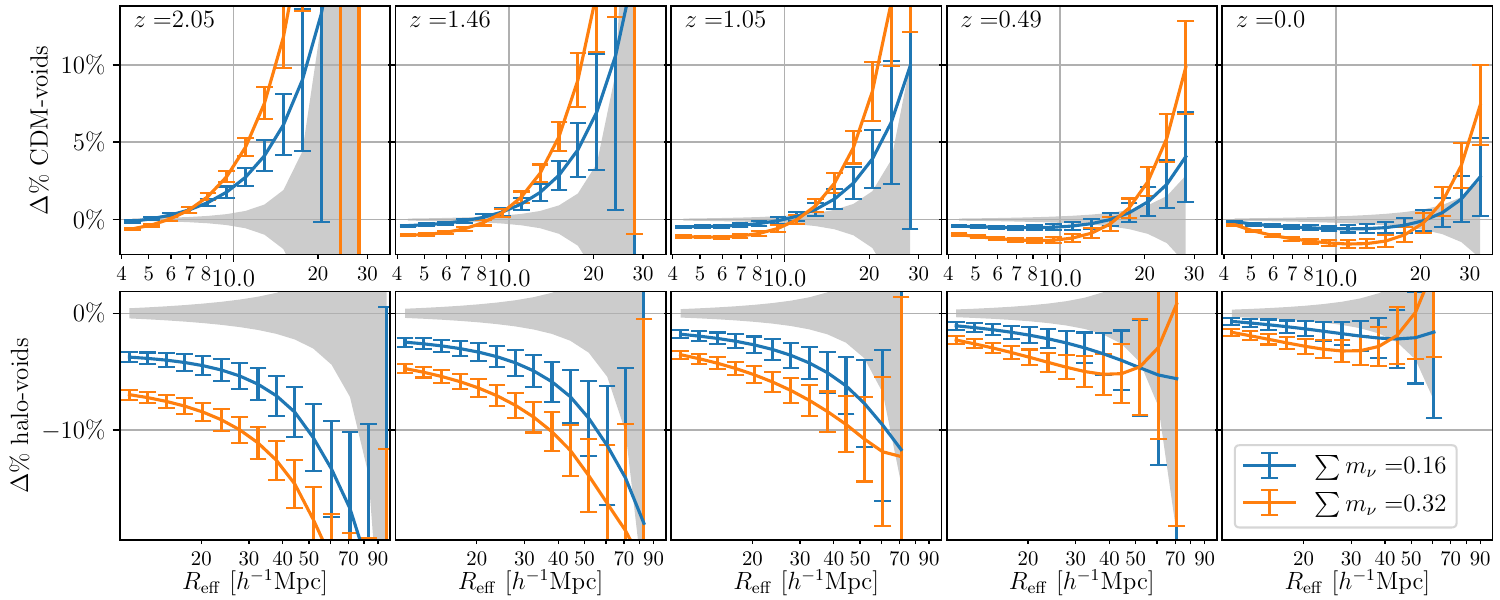}
\caption{VSF relative differences, with respect the \lcdm case, considering the cosmological constant case and neutrinos with total masses $\sum m_\nu=0.16$ (blue) and $0.32$ eV (orange), at $z=0$, 0.49, 1.05, 1.46. For each case the VSF is computed with the Sheth and Van de Waygaert model~\citep{SVdW} with the power spectrum normalised at $\sigma_8=0.83$, as discussed in the text. The upper panels show CDM-traced voids, the lower panels halo-traced voids for $M \geq 2.5 \times 10^{12} h^{-1} M_\odot$. The errorbars are Poissonian errors, the grey shaded areas show the Poissonian errors for the \lcdm case. }
\label{fig:rel_neutrinos_sig8}
\end{figure}

\subsection{Massive neutrinos case}
For the massive neutrinos case, given the $D(z)$ scale-dependence produced by free-streaming neutrino, we compute the linear power spectrum directly with CAMB\footnote{\url{https://camb.info/}}~\citep{Lewis_2000_CAMB}, fixing $\sigma_8(z=0)$ equal to \lcdm case, by changing the $A_{\rm s}$ value. Then, we compute the corresponding $\sigma(R)$ and analytical VSF in the presence of massive neutrinos. As for the DE case, we repeat the same procedure for both CDM- and halo-voids and show the result of our procedure in Fig.~\ref{fig:rel_neutrinos_sig8}. For the neutrino masses considered, we can see distinguishable effects on the VSF with respect to the massless case. This is due to the fact that massive neutrinos effects cannot be encapsulated in the amplitude normalisation alone, but they also impact the power spectrum shape and the (scale dependent) linear growth factor, as functions of redshift. The differences with respect to the massless case grow as the redshift increases.
Therefore, we conclude that, even when using cosmologies with matched $\sigma_8$ at $z=0$, the VSF is able to distinguish between different $\sum m_\nu$ values, as well as different DE-EoS and their combined effect. Moreover, in some cases, using $\sigma_8$-matched cosmologies, the difference between the models can even be larger than in the $A_{\rm s}$-matched case.

\subsection{Propagating Planck $A_s$($\sigma_8$) errors to the VSF}

In this subsection, we consider how the uncertainty on $A_{\rm s}$ constrained on CMB experiments propagates in $\sigma_8$ and in the void size function. The uncertainty on $A_{\rm s}$ obtained by Planck~\citep{planck_2018} assuming the \lcdm model, reflects in a relative error of $0.7\%$ on $\sigma_8$. If we assume that VSF can be described by a multiplicity function depending on $\sigma$ of the form
\begin{equation}
\frac{{\rm d} n(R)}{{\rm d} R}  =  \frac{f(\sigma)}{V(R)} \frac{{\rm d} \sigma}{{\rm d} R}\,,
\end{equation}
the uncertainty on $A_{\rm s}$ can be propagated as follow. The derivative with respect $A_{\rm s}$ can be written as
\begin{equation}
\frac{{\rm d} }{{\rm d} A_{\rm s}} = \frac{{\rm d} \sigma_8}{{\rm d} A_{\rm s}} \frac{{\rm d} }{{\rm d} \sigma_8} \,.
\end{equation}
When the shape of $\sigma(R)$ does not depend or weakly depends on the initial scalar amplitude, the variation of the VSF with respect to $A_{\rm s}$ can be written as
\begin{equation}
\frac{\rm d}{{\rm d} A_{\rm s}} \frac{{\rm d} n}{{\rm d} R} = \frac{1}{\sigma_8} \frac{{\rm d} \sigma_8}{{\rm d} A_{\rm s}} \left[  \frac{{\rm d} n}{{\rm d} R} + \sigma(R)  \frac{{\rm d}}{{\rm d} \sigma} \frac{{\rm d} n}{{\rm d} R} \right] \,.
\end{equation}
It follows that the uncertainty on $A_{\rm s}$ propagated to the VSF corresponds to
\begin{equation}\label{eq:VSF_As_err}
\Delta\% = \frac{\Delta \sigma_8}{\sigma_8} \left[  1 + \sigma(R) \left(\frac{{\rm d} n}{{\rm d} R}\right)^{-1} \frac{{\rm d}}{{\rm d} \sigma} \frac{{\rm d} n}{{\rm d} R}  \right]\,.
\end{equation}
\begin{figure}[t!]
\centering
\includegraphics[width=0.65\textwidth]{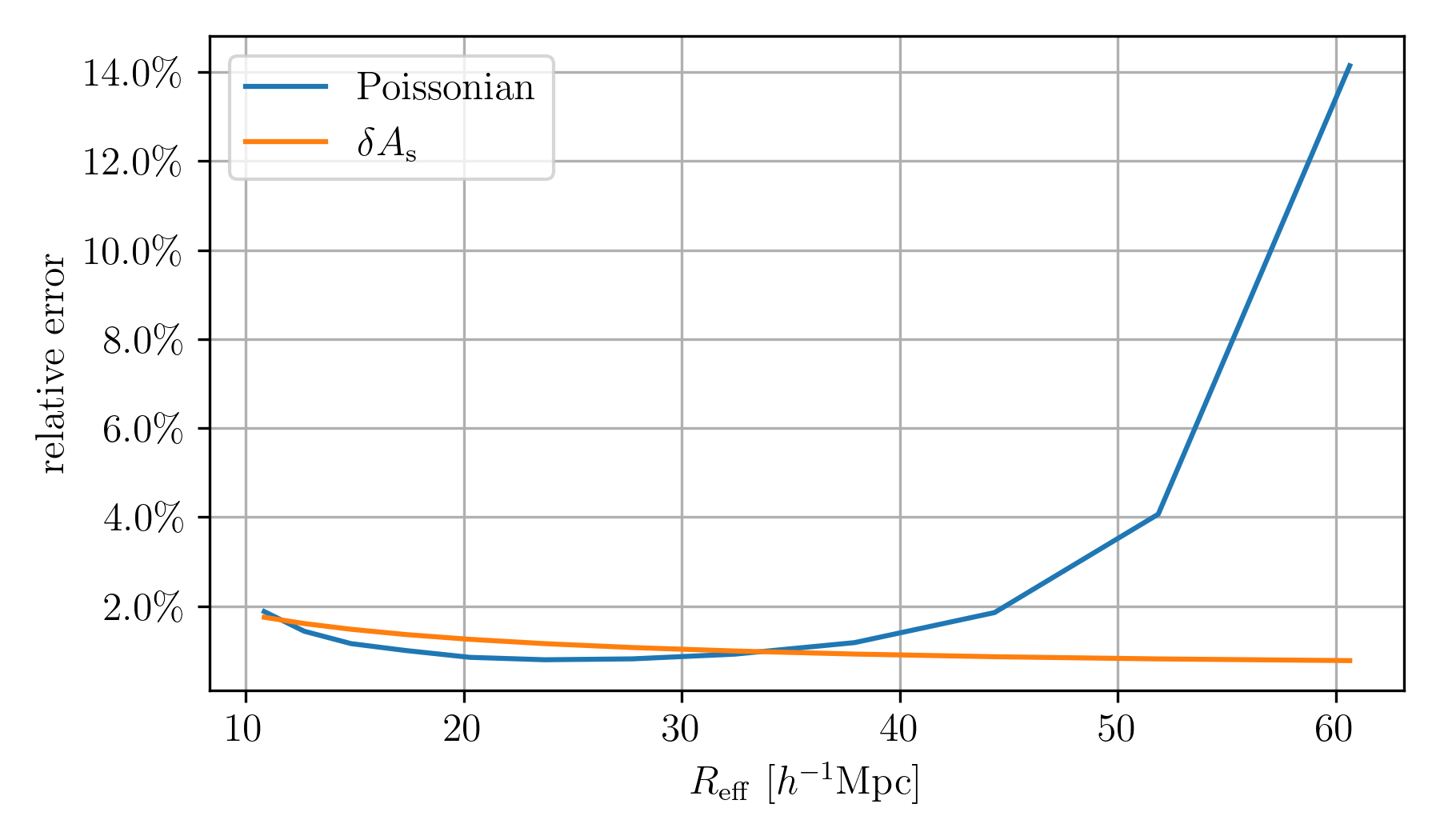}
\caption{VSF relative Poissonian error (blue) and uncertainty of $A_{\rm s}$ constrained by Planck~\citep{planck_2018} propagated to the VSF (orange, see text). The results refer to halo-traced voids for $M \geq 2.5 \times 10^{12} h^{-1} M_\odot$ at $z=0$ in the \lcdm case.}
\label{fig:As_err}
\end{figure}
Note that this result does not assume a specific VSF model, but only that the VSF can be modelled by a multiplicity function of $\sigma$. Fig.~\ref{fig:As_err} shows the results for halo-traced voids with $M \geq 2.5 \times 10^{12} h^{-1} M_\odot$ at $z=0$ in the \lcdm case. To compute the global derivative of the VSF with respect to $\sigma$ appearing in Eq.~\eqref{eq:VSF_As_err}, we numerically differentiated Eq.~\eqref{eq:SVdW} evaluated as described in Sec.~\ref{sec:sig8_DE} for halo-voids at slightly different $\sigma_8$ values. It can be noticed that the derived uncertainty is of the same order of magnitude as the Poisson error.

\end{document}